\documentclass[%
 reprint,
 floatfix,
 amsmath,amssymb,
 prstab,
]{revtex4-2}

\usepackage{graphicx}% Include figure files
\usepackage{dcolumn}% Align table columns on decimal point
\usepackage{bm}% bold math
\usepackage{subcaption}% subfigures
\usepackage{textcomp} % to get rid of warnings from gensymb
\usepackage{gensymb} % degree symbol

\usepackage{booktabs} % for toprule, midrule, bottomrule
\usepackage{multirow, units}
\usepackage{placeins}
\usepackage{textgreek}
\usepackage{lineno}

\begin{document}
\preprint{APS/123-QED}

\title{Fast Failures in the LHC and the future High Luminosity LHC}
\thanks{Supported by the HL-LHC project}

\author{B.~Lindstrom\textsuperscript{1,2},
P.~B\'elanger\textsuperscript{1},
L.~Bortot\textsuperscript{1}, R.~Denz\textsuperscript{1}, M.~Mentink\textsuperscript{1}, E.~Ravaioli\textsuperscript{1}, F.~Rodriguez~Mateos\textsuperscript{1}, R.~Schmidt\textsuperscript{1}, J.~Uythoven\textsuperscript{1}, M.~Valette\textsuperscript{1}, A.~Verweij\textsuperscript{1}, C.~Wiesner\textsuperscript{1}, D.~Wollmann\textsuperscript{1}, M.~Zerlauth\textsuperscript{1}\\
\textsuperscript{1}CERN, Geneva, Switzerland, \textsuperscript{2}Uppsala University, Uppsala, Sweden}

\date{\today}% It is always \today, today,

\begin{abstract}
An energy of $\unit[362]{MJ}$ is stored in each of the two LHC proton beams for nominal beam parameters. This will be further increased to about $\unit[700]{MJ}$ in the future High Luminosity LHC (HL-LHC) and uncontrolled beam losses represent a significant hazard for the integrity and safe operation of the machine. In this paper, a number of failure mechanisms that can lead to a fast increase of beam losses are analyzed. Most critical are failures in the magnet protection system, namely the quench heaters and a novel protection system called Coupling-Loss Induced Quench (CLIQ). An important outcome is that magnet protection has to be evaluated for its impact on the beam and designed accordingly. In particular, CLIQ, which is to protect the new HL-LHC triplet magnets, constitutes the fastest known failure in the LHC if triggered spuriously. A schematic change of CLIQ to mitigate the hazard is presented.

A loss of the Beam-Beam Kick due to the extraction of one beam is another source of beam losses with a fast onset. A significantly stronger impact is expected in the upcoming LHC Run III and HL-LHC as compared to the current LHC, mainly due to the increased bunch intensity. Its criticality and mitigation methods are discussed.

It is shown that symmetric quenches in the superconducting magnets for the final focusing triplet can have a significant impact on the beam on short timescales. The impact on the beam due to failures of the Beam-Beam Compensating Wires as well as coherent excitations by the transverse beam damper are also discussed.

\end{abstract}

\maketitle

\section{INTRODUCTION}

The stored beam energy in the LHC for the nominal beam parameters is $\unit[362]{MJ}$. This presents several challenges for the safe operation of the accelerator, for which a sophisticated interlock and beam dumping system has been implemented~\cite{schmidt_protection_2006}. This allows for the beam to be extracted safely when any kind of failure that risks leading to uncontrolled beam losses, and subsequent equipment damage, is detected. For machine protection purposes, it has been determined that quasi-instantaneous beam losses exceeding $\unit[1]{MJ}$ into the collimation system~\cite{LHCTDR_chapter18,assmann05chamonix,assmann06,bruce14_PRSTAB_sixtr,valentino17_PRSTAB}, designed to clean the bunch tails and to provide a passive protection against equipment damage, are considered critical. The exact damage limits vary on a case-by-case basis due to different loss rates, loss duration, impact factors and loss locations, and the mentioned limit has thus been determined with adequate margins.

The High Luminosity LHC (HL-LHC) project~\cite{HLTDRv01_chapter1} will provide several upgrades of equipment and beam parameters, in order to achieve about a factor eight higher peak luminosity in the ATLAS and CMS experiments. This will come with an increased stored beam energy, to about $\unit[700]{MJ}$, as well as increasing the beta functions, and thus the sensitivity of the beam to failures, at critical locations by up to about a factor four. This, together with new types of equipment introducing other potential failures, implies a significant challenge for machine protection.

A failure is any kind of unwanted equipment behavior or operational manipulation, and if it affects the beam, critical beam losses can result. The focus of this paper lies on \textit{fast failures} that lead to beam losses risking equipment damage within $\unit[10]{ms}$ from the onset of failure. These failures are the most critical and require reliable and fast failure detection and interlocking, as well as passive protection schemes. 

The paper starts with an explanation of the relevant parts of the HL-LHC upgrade, followed by an explanation of the main machine protection systems and the assumptions made on the bunch distribution, which is important for the beam loss estimates. Next, the main effects on the beam due to failures, orbit distortions and beta beating, are explained to better understand the results of the individual failure cases. The potential critical failures discussed in this paper then follow:

\begin{itemize}
    \item failures related to the magnet protection equipment, namely Quench Heaters (QH) and Coupling-Loss Induced Quench (CLIQ)
    \item a loss of the Beam-Beam Kick (BBK) due to the dumping of one beam
    \item failure of the long range Beam-Beam Compensating Wires (BBCW)
    \item coherent excitations by the transverse damper (ADT)
    \item fast-propagating symmetric quenches of the triplet magnets
\end{itemize}

Asynchronous beam dumps and injection kicker failures also constitute significant, fast, failures. Injection kicker failures have been studied in detail by other authors~\cite{bruning_impact_1999} and will not be discussed in this paper. The consequences of asynchronous beam dumps have also been studied in detail by other authors~\cite{wiesner_asynchronous_2018}, and will only be briefly discussed.

\section{High Luminosity LHC Upgrade}
The HL-LHC project, planned for completion in 2027, consists of an upgrade of the LHC with the aim of reaching about a factor eight larger instantaneous luminosity~\cite{HLTDRv01_chapter1}. The injector chain upgrade will provide the LHC with bunches of $2.2 \times 10^{11}$ protons per bunch and a normalized emittance of $\unit[2.5]{\mu m \cdot rad}$. The increased bunch intensity is the cause of the increased stored beam energy by about a factor of two, since the total number of bunch slots will remain the same as in the LHC era. The proton energy will be $\unit[7]{TeV}$.

Within the eight insertion regions (IRs) of the LHC, there are four interaction points (IPs): ATLAS (IP1), ALICE (IP2), CMS (IP5) and LHCb (IP8). The goal will be to squeeze the \textbeta\textsuperscript{*} in IPs 1 and 5 from the 2018 LHC value of $\unit[25]{cm}$ down to $\unit[15]{cm}$, implying larger beta functions in the final focusing triplets, with values up to $\unit[\sim22]{km}$.

Some of the main components of the upgrade are the ATLAS and CMS final focusing triplets, consisting of three superconducting quadrupole magnet assemblies, Q1, Q2 and Q3, with an increased aperture from the current $\unit[70]{mm}$ to $\unit[150]{mm}$ diameter and a decreased gradient from $\unit[200]{T/m}$ to $\unit[132.6]{T/m}$~\cite{HLTDRv10_chapter3}.
Each triplet quadrupole is split into two separate, equivalent, magnet halves and all magnets of a triplet are powered in a single circuit with three nested power converters~\cite{HLTDRv10_chapter6}.
The increased aperture will allow the beam size and separation to be larger inside the triplets, necessary for achieving the small \textbeta\textsuperscript{*}. The maximum orbit is increased from $\unit[6]{mm}$ in LHC (2017, 30 cm collision optics) to $\unit[17]{mm}$ in HL-LHC (v1.4~\cite{maria_high_2019}, 15 cm collision optics), with the beam separation being about a factor two larger. This gives approximately a factor two stronger fields at the beam position despite the lower gradient.

The separation and recombination dipoles (D1 and D2) of ATLAS and CMS, which are currently resistive magnets, will be replaced by superconducting magnets, to gain space for other equipment. Two main dipoles in the dispersion suppressors in cell 9 of the betatron collimation region in IR7 will also be replaced, each by a pair of $\unit[11]{T}$ dipole magnets based on Nb\textsubscript{3}Sn superconductor, hereinafter referred to as \textit{the} $\unit[11]{T}$ dipoles. This allows using shorter magnets, such that extra collimators required to improve the cleaning efficiency can be installed in-between the magnets of each pair. There are more magnets in the IRs that will be upgraded, but those are not critical in view of fast failures and are not considered in this paper. 

Other key components of the HL-LHC project are the crab cavities. Since the bunch intensity is increased and the \textbeta\textsuperscript{*} decreased, the crossing angle must be increased in order to limit the long-range beam-beam effects. However, this would decrease the effective overlap of colliding bunches in the IP, leading to a significant reduction of the luminosity. This geometric reduction factor is about $0.8$ in the current LHC, but would go down to $0.3$ in the HL-LHC era. To compensate for this, ATLAS and CMS will have two crab cavities installed on each side of the IP, for each beam (total of eight crab cavities). By applying a longitudinally modulated transverse kick, the bunches are effectively tilted leading to a better overlap in the collision point. While vital for achieving the luminosity goal, the crab cavities give strong transverse kicks on the beam and it must therefore be ensured that potential fast failures are appropriately interlocked. The implications of these fast failures for machine protection are presented in other studies~\cite{garcia_experiment_nodate,crab_quench,lindstrom_crab_2018,lindstrom_machine_2019}.

Two other types of equipment considered for HL-LHC are the Hollow Electron Lens (HEL), which was added during the writing of this paper, and the Beam-Beam Compensating Wires (BBCW), which remain an option. The HEL aims to use a hollow electron beam to suppress the halo of the main proton beam, whereas the BBCW are current-carrying wires with the purpose of compensating some of the beam-beam effects. The HEL is only discussed briefly in this paper, while the BBCW are analyzed for potential failure modes. 

Preceding the HL-LHC era of the LHC, the LHC in its current configuration will be in its third and last run, the so called \textit{LHC Run III}, from 2021 to 2024. Where relevant, failure cases are also considered for LHC Run III optics and beam parameters. These parameters are detailed where used.

\section{Machine Protection Systems}
The machine protection systems can be split into passive and active components~\cite{schmidt_protection_2006}. Active components detect abnormal beam conditions with the purpose of extracting the beams, whereas passive protection is designed to diffuse and absorb beam losses.

\subsection{Passive Protection}
Passive protection consists of collimators defining the limiting aperture of the LHC. In the collimation insertions, particles from the beam halo are captured that could quench magnets if lost in the superconducting sections of the machine. If the beam is disturbed, beam losses first occur at the collimators. Absorbers are placed in front of sensitive elements with the goal of reducing particle showers into them.

The LHC employs a three-stage collimation system~\cite{LHCTDR_chapter18,assmann05chamonix,assmann06,bruce14_PRSTAB_sixtr,valentino17_PRSTAB}, consisting of primary and secondary collimators, as well as tertiary absorbers for protecting the final focusing triplet magnets. Together, the collimators cover all the phase advances of the betatron motion in the machine. The collimator apertures are set in units of RMS beam size $\sigma$, assuming a certain normalized transverse emittance ($\unit[3.5]{\mu m \cdot rad}$ in LHC and $\unit[2.5]{\mu m \cdot rad}$ in HL-LHC). The corresponding values in millimeters thus depend on the local beta function and beam energy. The primary collimators have the tightest aperture of $\unit[6.7]{\sigma}$ (in HL-LHC), which is equivalent to about $\sim\unit[1.5]{mm}$ at $\unit[7]{TeV}$. These apertures are centered around the measured beam orbit and the full gap width of a collimator is twice the mentioned value. Some collimators have a single-sided jaw, such as the movable dump absorbers.

In this paper all values have been calculated using a normalized emittance of $\unit[2.5]{\mu m \cdot rad}$, since this is the design value of HL-LHC, and is also closer to the actual value in LHC Run II and Run III. A summary of the collimator gap settings for LHC Run II (2018 values) and HL-LHC can be seen in Table~\ref{tab:collSettings}. Run III settings are yet to be defined, but are likely to be similar to Run II. In this paper, they are thus assumed to be the same.

\begin{table}[!htb]
    \centering
    \caption{Nominal settings in units of RMS beam size $\sigma$ of selected collimator gaps in Run II and HLLHCv1.3 optics, assuming a normalized emittance of $\unit[2.5]{\mu m\cdot rad}$, in collision optics~\cite{HLTDRv10_chapter5}. The settings apply equally in horizontal, vertical and skew setups.}
    \begin{tabular}{lcc}
         Element & Run III & HL-LHC \\
         \toprule
         IR7 primary collimators (TCP)            & 5.9  & 6.7\\
         IR7 secondary collimators (TCS)           & 7.7  & 9.1\\
         IR6 dump absorbers (TCDQ)    & 8.6 & 10.1\\
         IP1/IP5 tertiary absorbers (TCT)    & 9.2 & 10.4\\
         Triplet aperture (IP1)   & 10.4 & 11.2\\
         Triplet aperture (IP5)   & 10.4 & 11.5\\
         \bottomrule
    \end{tabular}
    \label{tab:collSettings}
\end{table}

\subsection{Active Components}
The active components of the machine protection systems, such as the Beam Loss Monitors (BLM), feed into the beam interlock system, which subsequently activates the LHC beam dumping system to extract the beams (a \textit{beam dump}) when abnormal conditions are detected. The BLM system~\cite{BLMsystem} constitutes a general purpose active protection, consisting of approximately 3700 ionization chamber BLMs, distributed throughout the whole ring. This allows determining the location of any beam losses with a $\unit[40]{\mu s}$ temporal resolution (one LHC turn is $\unit[89]{\mu s}$), as well as acting on losses in $\unit[80]{\mu s}$ to trigger beam dumps for critical losses.

Normally, machine protection relies on detection of equipment failures and on dumping the beams before they are affected by the failure. However, for some failures, such as beam-macroparticle interactions (UFOs)~\cite{lindstrom_results_2018} or transverse beam instabilities, the BLMs are the only devices capable of detecting the failure and dumping the beams. 
The BLM system is designed for local protection against equipment damage due to beam losses, but since any failure critical for machine protection eventually causes beam losses at the primary collimators, the BLMs also provide a global protection of the machine~\cite{schmidt_protection_2006}.

A novel system to be installed in the LHC consists of two beam current change monitors, which are solely a global protection system measuring the change in stored beam current with a high precision better than $3\times10^{11}$ protons for single-turn ($\unit[89]{\mu s}$) integration windows. The interlock levels have been proposed at $3\times10^{11}$ protons for the single-turn integration window. This system will provide diverse redundancy to the existing BLMs~\cite{BCCM,BCCM_EDMS}.

\subsection{Reaction Time}
The time to detect a failure and initiate a beam dump is one of the key parameters when it comes to the protection of the machine.

After a failure has been confirmed by some system, it is communicated via the beam interlock system to the LHC beam dumping system, which can take up to $\unit[100]{\mu s}$. This system can then execute a beam dump after synchronizing the firing of the dump kicker magnets to the particle free abort gap (up to $\unit[89]{\mu s}$) and then extract the beam ($\unit[89]{\mu s}$). One can therefore only be certain that both beams are dumped after $\unit[278]{\mu s}$ or about three LHC turns. Furthermore, the failure must first be detected. The fastest detection systems in the current LHC are the fast magnet current change monitors ($\unit[\sim20]{\mu s}$) measuring the current of several non-superconducting magnet systems, and the BLMs ($\unit[\sim80]{\mu s}$). In the HL-LHC there will also be an addition of interlocking of the new crab cavities with detection times of $\sim\unit[2]{\mu s}$ using the low-level radio-frequency controller. In general, these systems detect an abnormal condition, and then confirm the abnormality over a certain period of time to not cause unnecessary beam dumps on temporary spikes or noise. This evaluation time depends strongly on the observed signal and design and can take up to a few LHC turns. Therefore, it is required from a machine protection point of view that any fast failure does not cause any critical beam losses within less than ten LHC turns, to allow sufficient time for failure detection and safe extraction of the beams. Hereinafter, \textit{LHC turn} is referred to as \textit{turn}.

\section{Transverse Particle Distribution}
The transverse bunch profiles in accelerators are often modelled by a Gaussian distribution. However, as has been seen in the LHC through beam losses from collimator scraping measurements~\cite{Arek} and van der Meer scans~\cite{VDMdistribution}, the LHC bunches tend to have overpopulated transverse tails as compared to a true Gaussian distribution. This has a significant impact on the criticality for machine protection, as it increases the energy stored in the tails and therefore the energy deposited into the aperture for small beam perturbations. For loss estimates in this paper, the projection of the bunch distribution onto the horizontal and vertical axes is modeled as the addition of two concentric Gaussian distributions, one containing $\unit[80]{\%}$ of the particles, and one containing $\unit[20]{\%}$ of the particles with an RMS width a factor two larger. This \textit{heavy tailed} distribution approximates the collimator scraping measurements.

\subsection{Beam Orbit Excursion Limit}
\begin{figure}[!htb]
   \centering
   \includegraphics*[width=0.48\textwidth]{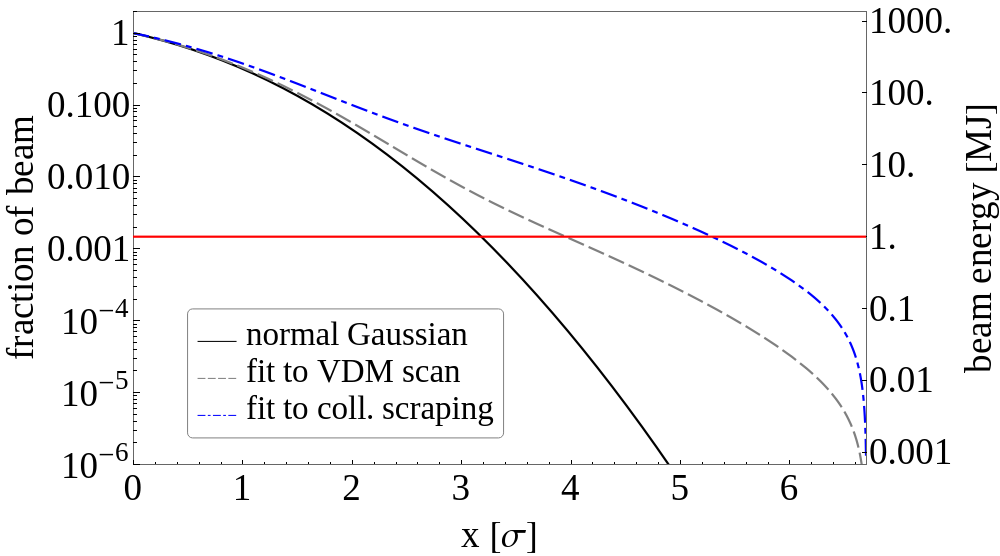}
   \caption{Comparison of the fraction of particles outside a certain transverse position, when projected onto the horizontal or vertical axis, for three different bunch distributions: a normal Gaussian, one based on the van der Meer (VDM) scans and one based on collimator scraping measurements. The energy scale shows the corresponding energy contained in that fraction of the nominal beam. The $\unit[1]{MJ}$ limit is shown in red.}
   \label{fig:bunchDistribution}
\end{figure}

The beam size \textsigma, in Fig.~\ref{fig:bunchDistribution} and elsewhere in this paper, refers to the assumed RMS size of a Gaussian beam given a normalized emittance of $\unit[2.5]{\mu m \cdot rad}$. The RMS width of the heavy tailed distribution is about a factor 1.3 larger than this. 

A comparison of the fraction of particles outside a certain transverse position, when projected onto the horizontal or vertical axis, for different bunch distributions is shown in Fig.~\ref{fig:bunchDistribution}. Beam losses exceeding $\unit[1]{MJ}$ have been defined as critical and should be avoided. This limit is shown in red assuming a full HL-LHC beam of 2748 bunches with a bunch intensity of $\unit[2.2\times10^{11}]{p^+}$. Beam orbit excursions larger than $\unit[1.5]{\sigma}$ are thus considered critical using the collimator scraping data. For LHC Run III beam parameters, this limit is at $\unit[1]{\sigma}$ due to the tighter primary collimator gap settings (see Table~\ref{tab:collSettings}).

\subsection{Impact of Hollow Electron Lens}
Hollow Electron Lenses (HEL)~\cite{HEL2,HEL1} have recently been added to the HL-LHC baseline. The HEL works by superimposing a low-energy electron beam concentrically around the main proton beam, such that it only overlaps with the halo of the proton beam. This increases the diffusion speed for particles with a large transverse action in the main beam, effectively scattering them into the collimators. The bunches consequently get a smaller halo density, such that small orbit perturbations would lead to significantly smaller beam losses. This could be beneficial for any kind of failure, as it increases the time from \textit{onset} of failure until critical losses occur.

However, one vital parameter for machine protection is the time between \textit{detection} of a failure and beam dump. With the depleted halos the transverse particle distribution in a bunch could be narrower, such that for large orbit perturbations, the resulting beam losses would rise faster from undetectable to critical. This would impact the time margins for the BLM system as well as the new beam current change monitor and affect their efficiency in case of a fast failure. 

While the HEL is in general beneficial, to ensure the reliability of the machine protection, one requirement would be that there are untouched \textit{witness bunches} that have their halos intact. In combination with adjusted interlock levels for the BLMs and the beam current change monitor, these will act as an early warning, such that the time between detection of a failure and beam dump remains sufficiently long. 

Detailed studies on the impact of the HEL on the failure cases discussed in this paper and the use of witness bunches in combination with adjusted interlock thresholds are currently ongoing, but outside the scope of this paper. 

\section{Effect on beam due to fast failures}\label{orbitEffects}
There are two major effects that can be critical for the short-term behavior of the beam:
\begin{itemize}
\item The beam offset which is caused by dipolar field components, as well as any higher order components whenever the beam is not centered in the magnet. The latter is the case in the triplet magnets, where the nominal orbit is offset by $\unit[>10]{mm}$ as compared to the magnetic center.
\item The beta beating resulting from a change of the quadrupolar focusing gradient, as well as higher order components, which can directly induce losses, as well as changing the hierarchy of the collimation system, affecting the protection of the aperture.
\end{itemize}

In general, the former is the most critical consequence as the subsequent beam losses are due to the beam being displaced, leading to relatively fast losses. Beta beating can lead to increased losses due to the beam size being modulated throughout the machine, changing the effective gap settings of the collimators, and potentially the collimator hierarchy. This kind of losses are more diluted and would be less likely to cause damage before the BLM system acts to dump the beams. Nevertheless, strong beta beating can induce losses in the primary collimators that could be critical. 

Except when stated otherwise, orbit excursions and kicks in this paper refer to the linear action in phase space, and are normalized to the transverse beam size in units of one standard deviation under the assumption of normally distributed transverse beam profiles. The \textit{heavy tailed} distribution above is only used for loss estimates and for determining the orbit excursion threshold.

\begin{figure}[!htb]
   \centering
   \includegraphics*[width=0.48\textwidth]{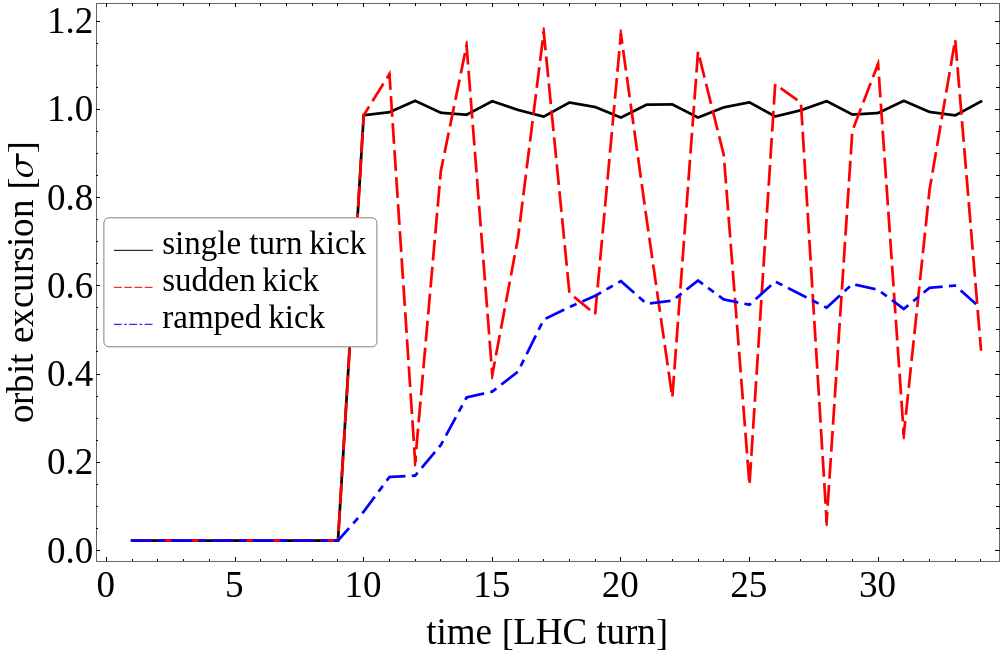}
   \caption{Comparison of the orbit excursion over time for three different types of kicks. The kick strength in all cases is $\unit[1]{\sigma}$. For the solid line, the kick was present only on one single turn. For the dashed-dotted line, the kick was ramped over ten turns and then remained constant, while for the dashed line the kick was switched to maximum strength in one turn and then remained constant.}
   \label{fig:orbitBehavior}
\end{figure}

The orbit perturbations due to fast failures can have different characteristics, depending on how fast the onset of the kick is, its amplitude, and whether or not it is transient or constant. The normalized beam orbit excursion over time, as simulated for three different types of vertical kicks in the LHC, is shown in Fig.~\ref{fig:orbitBehavior}. Note that the behavior for horizontal kicks would be the same.

The solid line shows the beam orbit change over time if a kick of $\unit[1]{\sigma}$ is applied on one turn, and then immediately switched off. The beam will then oscillate with the betatron tune around the original closed orbit, but the \textit{orbit excursion} remains approximately constant at $\unit[1]{\sigma}$.

The dashed-dotted line also shows a kick of $\unit[1]{\sigma}$, but it is slowly ramped up to full strength where it remains constant. The beam will then shift to a new closed orbit, with minimal oscillations around it. The difference between the initial and this new closed orbit is a factor $2 \sin[\pi Q]$ smaller than the applied kick strength, with Q being the betatron tune, $60.3149$ in this case. The triplet quench and the Beam-Beam Compensating Wires discussed below approximately follow this behavior.

The dashed line shows the same kick, but it is suddenly ramped to full strength in one turn. The beam will then oscillate with the tune around the new closed orbit, with a maximum amplitude a factor two larger than the difference between the new and the initial closed orbits. Examples approximately following this behavior are the quench heaters and a loss of the Beam-Beam Kick, which are discussed further down. 

\section{Impact of Magnet Protection Devices on the Beam} \label{magnetProtection}
The energy stored in the final focusing triplet magnets (Q1, Q2 and Q3) around ATLAS and CMS will increase significantly for HL-LHC, up to a maximum of $\unit[8.37]{MJ}$ in the Q2 magnets~\cite{HLTDRv01_chapter3}, making it vital to ensure that the energy during a quench is dissipated throughout a large volume of the magnet. As for other superconducting magnets in the LHC, this is done by heating the magnet as soon as the quench is detected. There are two technologies, both part of the HL-LHC baseline~\cite{HLTDRv01_chapter7}, to attain this heating in the triplet magnets, quench heaters (QH) and the novel Coupling Loss Induced Quench (CLIQ) system. 

Quench heaters constitute the core of the magnet protection system of the current LHC. The QHs consist of resistive strips that are attached onto the magnet coil. When a quench is detected, a capacitor is discharged into these strips, heating them. The heat propagates into the magnet, making it quench from the outside inwards. The time it takes to reach maximum current in the QHs of the LHC main dipole magnets has been measured as $\sim \unit[30]{\mu s}$ and is estimated to be similar in the new HL-LHC magnets. This means that they can be considered instantaneous compared to a single turn of $\unit[89]{\mu s}$. 

CLIQ is a new system that functions by discharging a capacitor bank directly through dedicated current leads connected to the magnet coils~\cite{emmanuele}. The oscillatory nature of this current, reaching an amplitude of $\unit[1500]{A}$ in as little as $\unit[13]{ms}$, induces inter-filament and inter-strand coupling currents, heating the copper matrix surrounding the superconducting filaments. CLIQ is to be installed in the final focusing triplet magnets around ATLAS and CMS, where it is necessary for reducing the hot-spot temperature of the new HL-LHC Nb\textsubscript{3}Sn triplet magnets. 

In normal operation, i.e. when a quench is detected, it is required that the beam is extracted before the magnet protection system is activated, since both QHs and CLIQ induce magnetic fields in the beam region. These fields are capable of kicking the beam giving rise to critical beam losses on short timescales, as analyzed in detail below. Their effect on the beam during abnormal behavior, i.e. when there is a spurious discharge not triggered by a quench, is discussed in this section. 

The risk of common-mode failures triggering multiple quench protection systems simultaneously is deemed small enough to be neglected. For all failures, it is assumed that only one unit can trigger spuriously, meaning one circuit of two QH strips (see Fig.~\ref{fig:QHschematic}) or one CLIQ unit (see Fig.~\ref{fig:CLIQcircuit}). For QHs, the impact of a normal discharge on the beam, in case it is not extracted before firing, is also discussed. 

\subsection{Optics}
The optics used in this section are \textit{HLLHCv1.4}~\cite{maria_high_2019}, round collision optics with a \textbeta\textsuperscript{*} of $\unit[15]{cm}$, and \textit{flat optics}. Flat optics are an alternative in case of problems with the use of Crab Cavities. There the beam is strongly focused in only one plane at the IPs, giving a \textbeta\textsuperscript{*} of $\unit[7.5]{cm}$ in the crossing plane and $\unit[30]{cm}$ in the other plane. This leads to roughly a factor two larger beta function in the off-crossing plane at the triplet magnets, and conversely a factor two smaller beta function in the crossing plane, as compared to the round optics.
Since this is currently a backup to the baseline optics, it is important to study the effects of failure cases for both types of optics.

\subsection{Quench Heaters}
During beam operations in 2016, at the top energy of $\unit[6.5]{TeV}$, an oscillation of the beam with a few tens of micrometers ($\unit[0.1]{\sigma}$, normalized to the beam size) was observed preceding a quench-induced beam dump~\cite{valette_impact_2018}. The source of this oscillation was linked to the firing of the QHs in a main dipole magnet. The quench protection system initiates a beam dump as well as triggering the magnet protection as it detects a quench. Due to delays inherent to the design, the actual dumping of the beam could be delayed by up to $\unit[5]{ms}$ as compared to the triggering of the QHs.

Following this, and in particular because of the fast rise time of the QH current of $\unit[\sim30]{\mu s}$, it was clear that magnet protection equipment must be evaluated for its impact on the beam. 

\begin{figure}[!htb]
   \centering
   \includegraphics*[width=0.48\textwidth]{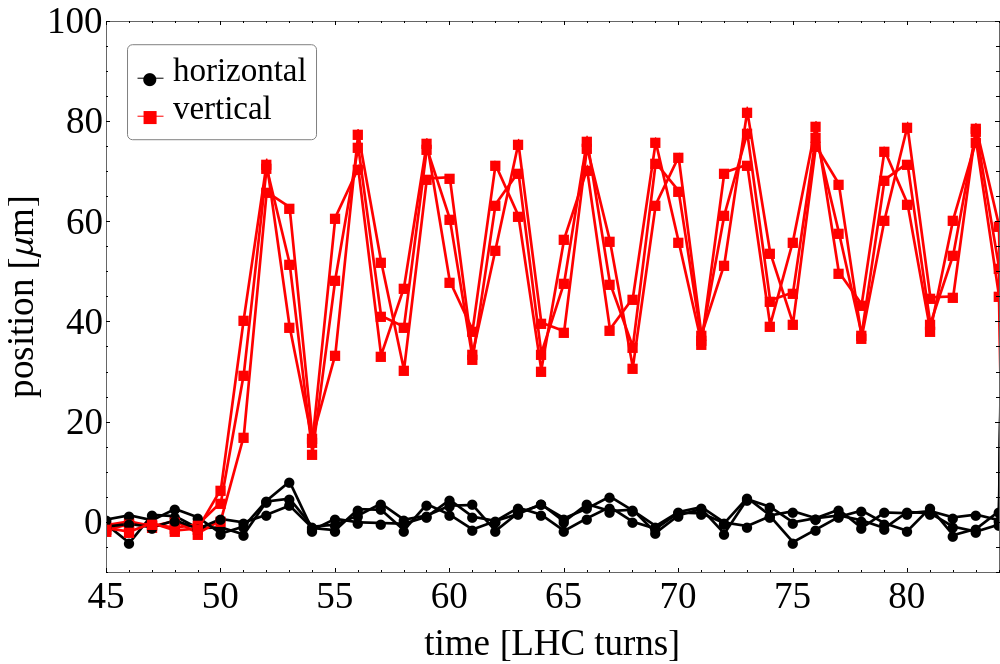}
   \caption{Orbit offset measured by a beam position monitor in IR4, for three bunches after the discharge of all the QH circuits in one normal dipole magnet. The beam energy was $\unit[3.46]{TeV}$.}
   \label{fig:QHoffset}
\end{figure}

Figure~\ref{fig:QHoffset} shows the orbit excursion of three bunches circulating in the LHC during an experimental firing of the QHs in 2018, at a beam energy of $\unit[3.46]{TeV}$. The amplitude was $\unit[\sim80]{\mu m}$. The kick is only in the vertical plane since the QH polarities are such that they are positive in one half of the magnet and negative in the other half, with a symmetry axis parallel to the horizontal axis.

The normalized kick during the experiment corresponded to $\unit[0.44]{\sigma}$, or an average magnetic field of $\unit[795]{\mu T}$ over the effective magnetic length of the dipole magnet~\cite{wiesner_lhc_nodate}.

The kick is small and of little concern for machine protection in the current LHC. However, in the HL-LHC era, the beta functions in the final focusing magnets will increase, and there will be an addition of QH circuits in several of the critical magnets~\cite{HLTDRv01_chapter7}, such as the final focusing triplets, separation (D1) and recombination (D2) dipoles and the $\unit[11]{T}$ dipoles.

The quench protection system for the new HL-LHC magnets and circuits will ensure that a beam dump is issued before a trigger is sent to the QHs and CLIQ. Nevertheless, a spurious firing of a single QH or CLIQ unit cannot be excluded. Therefore, this failure case has to be studied and its impact limited, such that unacceptable beam losses are avoided.

\subsubsection{Connection scheme}
In general, connecting two QH strips on opposite sides of the magnet will increase the impact on the beam, due to the relatively strong dipolar field induced in the beam region. Instead, connecting two strips next to each other, the contribution to the dipolar field in the beam region will be suppressed and the impact on the beam will be minimized. The former is the connection type in the LHC main dipole magnets, whereas the latter is the endorsed connection scheme for the D2 and the $\unit[11]{T}$ dipole magnets in the HL-LHC era. 

\begin{figure}
  \begin{subfigure}{0.48\textwidth}
    \centering\includegraphics[width=0.85\textwidth]{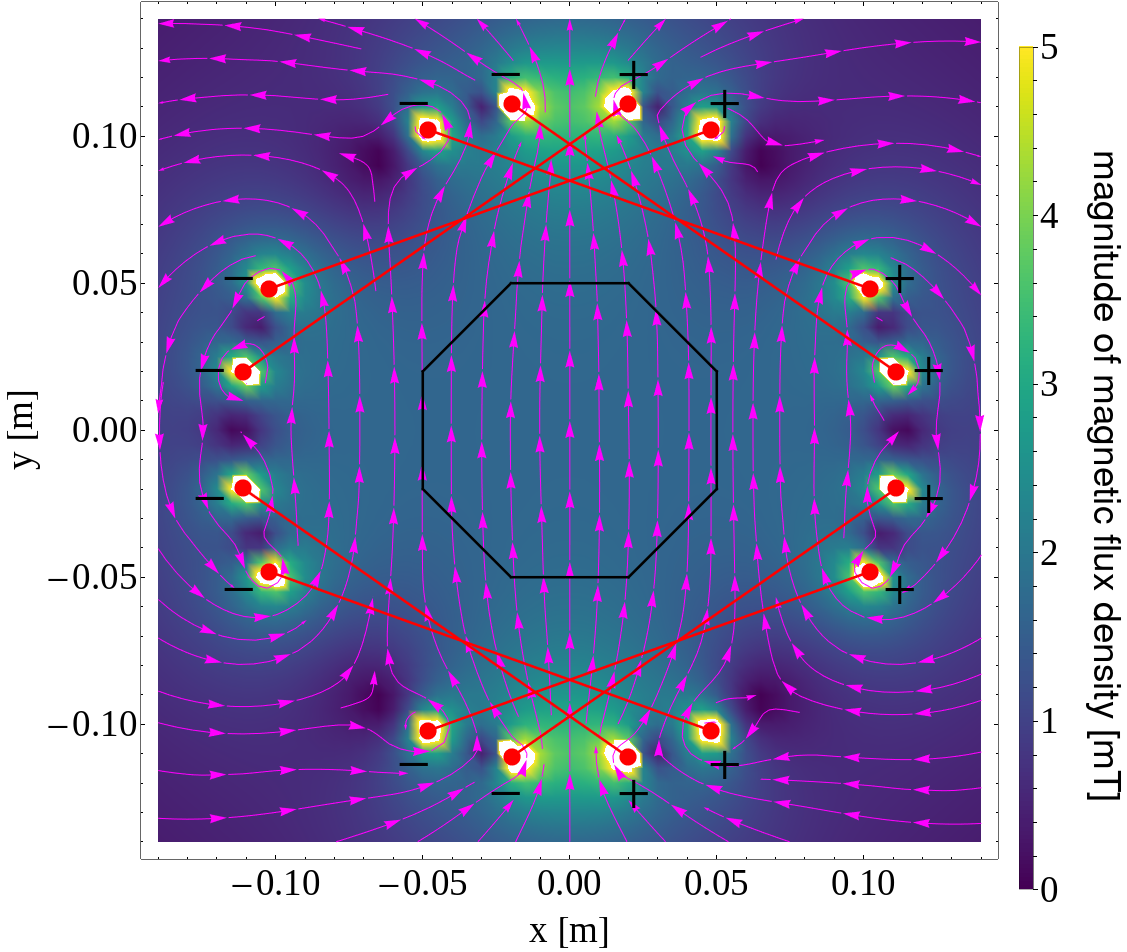}
    \caption{Triplet quadrupoles (Q1 beam screen)}
  \end{subfigure}
  \begin{subfigure}{0.48\textwidth}
    \centering\includegraphics[width=0.85\textwidth]{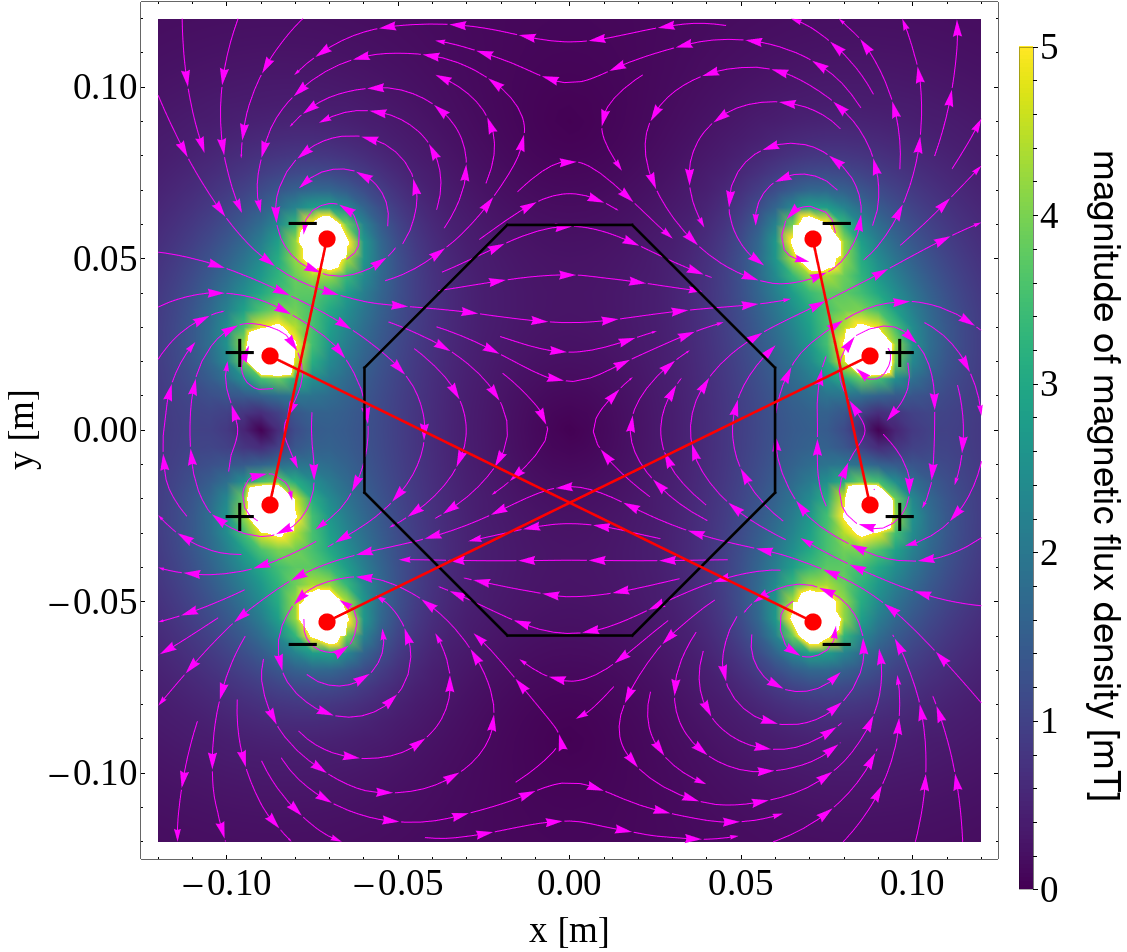}
    \caption{D1 separation dipole}
  \end{subfigure}
  \begin{subfigure}{0.48\textwidth}
    \centering\includegraphics[width=0.85\textwidth]{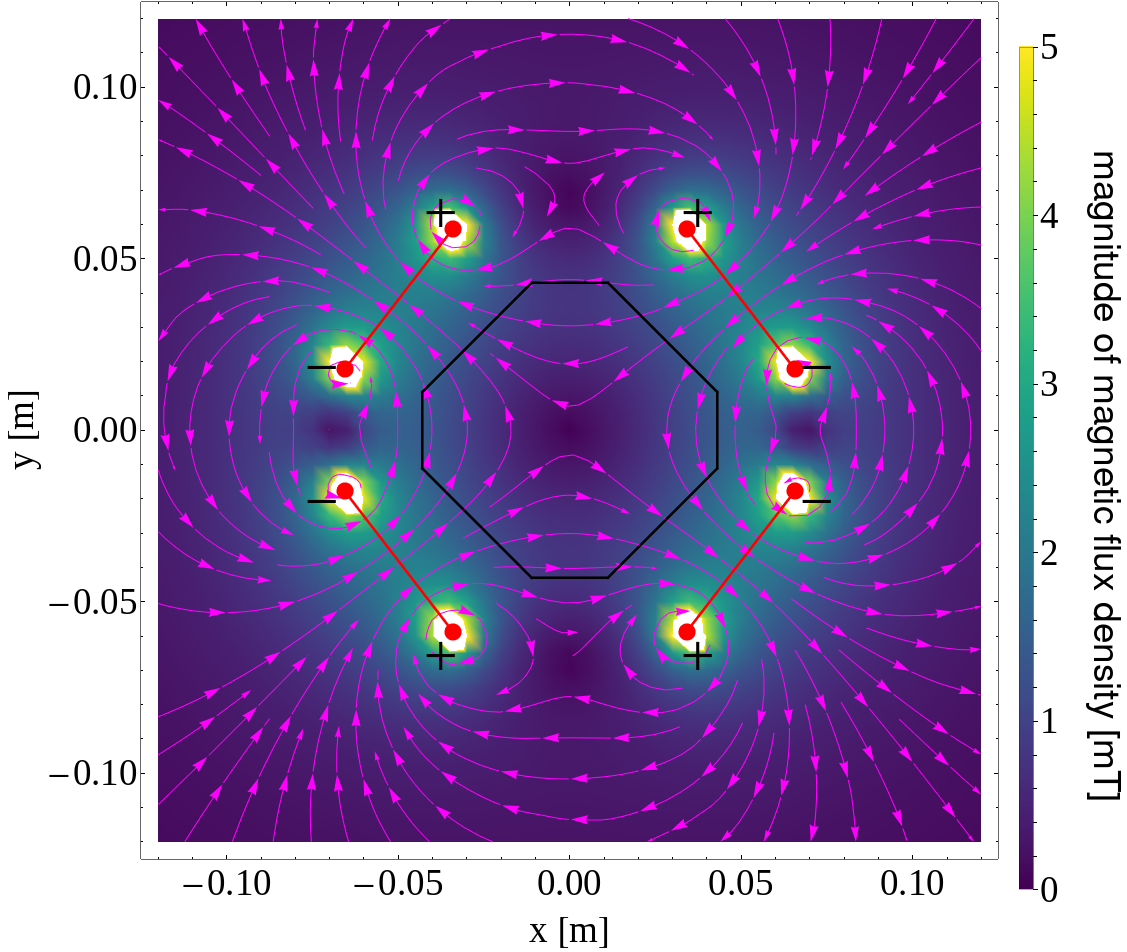}
    \caption{D2 recombination dipole}
  \end{subfigure}
  \caption{Magnetic fields during nominal QH discharges in selected HL-LHC magnets as calculated with a Biot-Savart model. The red points show the position of the QHs and the red lines show which QHs are connected to each other. The QH polarities are indicated by black + and -. The beam screen is shown for reference, but not used for the calculations.}
  \label{fig:QHschematic}
\end{figure}

The connection schemes, including the polarities of the QH strips, for the D1 and D2 dipoles, as well as the triplet quadrupoles, can be seen in Fig.~\ref{fig:QHschematic}. These plots also show the total magnetic field induced by a discharge in all the QH circuits. The D2 scheme, which is also applied for the $\unit[11]{T}$ dipoles, is the ideal case with respect to its impact on the circulating beam, without any connections across the magnet. The polarities are also distributed with equal negative and positive values on both sides of the magnets, giving but a small kick in the beam region.

For the D1, the distribution of QH polarities is the same as for the D2, meaning that a nominal discharge would give a quadrupolar field. However, the connection scheme is different, with two circuits going across the magnet. This connection scheme means that all quadrants of the magnet will be at least partially protected, in case one QH circuit malfunctions. However, for a spurious QH discharge, the effect on the beam could be significantly stronger than for the D2.

In case of the HL-LHC triplets, the focus was put on reducing the effect of a single QH circuit firing spuriously on the circulating beam~\cite{valette_impact_2018}. QH strips on neighboring poles are connected together, which partially mitigates the effect on the beam. However, the QH polarities are such that a strong dipolar field is produced, when all QHs are discharged. This can be accepted, as the quench detection system, in general, will ensure that the beams are dumped before sending a discharge trigger to the QH, and a simultaneous erratic of multiple QH circuits is excluded.

\subsubsection{QH kick}

The QHs, being $\unit[40]{mm}$ wide strips attached to the full length of the magnet coil and located at a radial distance of about a hundred millimeters from the beam~\cite{izquierdo_bermudez_overview_2018}, can be modeled as thin, infinitely long, current-carrying wires. At nominal current in the magnets the iron yoke is approximately saturated. The field induced by the QHs in the magnetostatic solution can thus be calculated by the Biot-Savart law. As there are usually two QH strips powered in series (one for the return current), the magnetic field seen by the beam is the superposition of the fields of at least two QH strips.

The infinitesimal kick $d\alpha$, given in units of beam RMS size ($\sigma$), that the beam receives can be calculated as follows,

\begin{equation}
d\alpha = \text{tan} \Bigg[\frac{B ds}{B\rho} \Bigg]\sqrt{\frac{\beta(s)}{\epsilon_g}}\approx \frac{B ds}{B\rho}\sqrt{\frac{\beta(s)}{\epsilon_g}}
\label{eq:kickIntegration}
\end{equation}

where $B$ is the magnetic field orthogonal to the considered kick, $B\rho$ the magnetic rigidity of the beam, $\beta(s)$ the beta function at position $s$, $ds$ an infinitesimal length along the magnet and $\epsilon_g$ the geometric emittance. The tangent function can be removed following the small angle approximation, since all considered kicks are on the order of microradians. Since the QHs reach their peak current within one turn, their effect was calculated by integrating the above equation during the peak of the discharge. As explained in section~\ref{orbitEffects}, for fast kicks that remain over multiple turns such as the QHs, the beam will oscillate around its new closed orbit. The difference between the new and the nominal orbits is a factor $2\sin[\pi Q]$ smaller than the applied kick, but the beam oscillation amplitude is up to a factor two larger than the difference between the orbits.

For the calculations, the average beta function and beam orbit throughout the magnet was used. This introduces a small error due to the variation of the beta function and orbit throughout the magnet. However, for the triplet magnets, having the largest variation, the error of the integrated kick is less than $\unit[1]{\%}$.

\subsubsection{Beam Screen}
The beam screen, present in all superconducting machine sections of the LHC, is expected to shield the beam region from fast transients in magnetic fields, effectively acting as a low-pass filter. It should, thus, play a role for both QH and CLIQ discharges, delaying their effects on the beam. The attenuation solely due to the beam screen for frequencies at $\unit[1]{kHz}$ has been calculated to be $\unit[15]{dB}$~\cite{morrone_magnetic_2019}. These calculations were based on periodically varying the current in a magnet, whereas the QH field rises in $\sim\unit[40]{\mu s}$ and then decays with a time constant of $\sim\unit[10]{ms}$. Measurements~\cite{wiesner_lhc_nodate} of the kick seen by the beam due to QH discharges in the LHC, imply a significantly weaker shielding than expected from~\cite{morrone_magnetic_2019}. In the measurements, the field seen by the beam reached $75\%$ of the externally applied field in two to three turns, followed by a slower rise of about 20 turns to reach the maximum. This discrepancy is as of the writing of this paper not understood. To be conservative for machine protection, all the results in this paper thus neglect the effect of the beam screen, such that the QH and CLIQ discharges have an immediate impact on the beam.

\subsubsection{Results}
The QH effect on the beam was calculated using \textit{HLLHCv1.4}~\cite{maria_high_2019} round and flat optics for Beam~1, taking the nominal orbits into account. For each magnet, the worst QH circuit was considered in the spuriously triggered case, and for each type of magnet, only the worst case is included in the results presented in Table~\ref{tab:QHresults}. For completeness, the QH currents are also noted in the table.

For nominal discharges, only the baseline round optics are shown. The beams are by design to be dumped before the QHs fire when correctly triggered. Nevertheless, if the beams were to remain in the machine, the kicks are relatively small for D1, D2 and the $\unit[11]{T}$ dipoles, with a maximum of $\unit[0.33]{\sigma}$ in the separation dipole (D1). For the triplet quadrupoles, the kick from nominal discharges is significantly worse, with up to $\unit[8.2]{\sigma}$ in one of the Q2 magnets. Given that all QH circuits in the full triplet (Q1, Q2 and Q3) are normally to be fired together, this would be an unacceptable failure case. For flat optics, given the magnitude of the kicks, the change of beta function by around a factor two is not enough to significantly reduce the normalized kicks, and the same conclusions thus hold.

For spurious discharges of single QH circuits, the $\unit[11]{T}$ dipole QHs still give a negligible kick, whereas in the D2 a vertical kick of up to $\unit[0.3]{\sigma}$ is generated. The D1 is significantly worse at $\unit[1.4]{\sigma}$, due to the QH circuit that goes across the magnet (see Fig.~\ref{fig:QHschematic}). The triplet quadrupoles also give significant kicks, mainly due to the large beta functions of up to around $\unit[22]{km}$. The worst kicks here are seen for one of the QH circuits in the Q2 magnets left of IPs 1 and 5 (as seen from the center of the LHC)%(MQXFB.B2L1, MQXFB.B2L5)
, reaching up to $\unit[1.2]{\sigma}$ horizontally. The vertical kick in these magnets is small, at $0.13$ and $\unit[0.33]{\sigma}$ respectively.

The reason for the triplets left of the IP having the largest kicks is that the magnetic field produced by the QHs is stronger vertically, for all circuits. The resulting horizontal kick combined with the largest horizontal beta functions in these Q2 magnets, provides the largest kicks. On the right side of the IP, the horizontal kick per magnet length is stronger in the Q3 magnets, but due to their shorter lengths ($4.15$ vs $\unit[7.15]{m}$), their integrated kicks are still smaller than the Q2 magnets on the left of the IP. This applies to both IPs 1 and 5, and the difference between them is small. 

For Beam~2 the optics are reversed, such that the largest horizontal beta function, and consequently also the largest kick, in the Q2 magnets is on the right side of the IP. One could imagine mitigating the kick by setting the QH polarities such that the field is aligned with the largest beta function, but since both beams share the same aperture in the triplet magnets, this would increase the kick in the other beam.

With flat optics the kicks are stronger due to the factor two larger beta functions in the off-crossing plane, reaching up to $\unit[1.7]{\sigma}$ in the Q2 left of IP1%(MQXFB.B2L1)
, and $\unit[1.8]{\sigma}$ in the worst D1 dipole. For the D2 dipoles, the increase is small, and the maximum kick kept at $\unit[0.43]{\sigma}$.

The beta beating due to the quadrupolar field components of below $\unit[8]{mT/m}$, is up to $\sim 1\%$, putting it well within the LHC design tolerances of $10\%$.

\begin{table}[tb]
    \centering
    \caption{QH results: kick strength of worst case single QH circuits as well as nominal firing of all circuits. Each triplet quadrupole is split in equal halves, designated by a and b, with a being closer to the IP.}
    \begin{tabular}{lccccc|c}
            & & \multicolumn{4}{c}{\textbf{single QH circuit}} & \multicolumn{1}{c}{\textbf{all circuits}} \\ 
            & & \multicolumn{2}{c}{Round} & \multicolumn{2}{c|}{Flat} & Round \\
        & QH current [A] & [$\sigma_x$] & [$\sigma_y]$ & [$\sigma_x$] & [$\sigma_y]$ & [$\sigma_r$] \\
        \toprule
        \multicolumn{6}{l|}{worst triplet magnets} \\
        Q1a & 200 & 0.33 & 0.12 & 0.46 & 0.09 & 2.31 \\
        Q1b & 200 & 0.44 & 0.14 & 0.61 & 0.09 & 3.03 \\
        Q2a & 200 & \textbf{1.17} & 0.10 & \textbf{1.59} & 0.16 & \textbf{7.87} \\ 
        Q2b & 200 & \textbf{1.23} & 0.13 & \textbf{1.70} & 0.22 & \textbf{8.25} \\ 
        Q3a & 200 & 0.69 & 0.09 & 0.97 & 0.16 & 4.64 \\
        Q3b & 200 & 0.76 & 0.07 & \textbf{1.04} & 0.14 & 5.04 \\
        \midrule
        \multicolumn{6}{l|}{worst separation dipole} \\     
        D1 & 168 & \textbf{1.38} & 0.39 & \textbf{1.78} & 0.26 & 0.33 \\
        \midrule
        \multicolumn{6}{l|}{worst recombination dipole} \\     
        D2 & 122 & 0.19 & 0.32 & 0.14 & 0.43 & 0.04 \\
        \midrule
        \multicolumn{6}{l|}{worst $\unit[11]{T}$ dipole} \\
        MBH & 150 & 0.01 & 0.03 & 0.01 & 0.03 & $<10^{-7}$ \\
        \bottomrule
    \end{tabular}
    \label{tab:QHresults}
\end{table}

\subsubsection{QH Conclusions}

Quench Heaters firing with beam in the machine constitutes one of the fastest failures, with a sub-turn onset. For the critical cases studied in this paper, that is, QHs in the final focusing triplet quadrupoles, and the separation (D1), recombination (D2), and $\unit[11]{T}$ dipoles, it will be ensured that the quench protection system does not trigger a QH to fire before the beams are completely extracted. Nevertheless, the QH circuits have been optimized to produce as small kicks on the beam as possible, that is, the dipolar field that they produce has been minimized as much as circuit limitations allow. For the dipole magnets, quadrupolar fields are produced and the kick on the beam only depends on a small nominal beam offset. For the triplet magnets, this was not possible, and a significant vertical dipolar field is still created.

It is excluded due to low probability that all QH circuits fire without being triggered, but this can not be ensured for spurious triggering of single circuits. For the D2 and the $\unit[11]{T}$ dipoles, this is not an issue, as each QH circuit lies solely in one quadrant of the magnet cross section. However, for the D1 and the triplet magnets, the QH circuits are connected across the cross section of the magnet, producing strong dipolar fields. The worst case within the triplets are the Q2 magnets, and they are of similar strength to the D1, with kicks of up to $\unit[1.38]{\sigma}$ in round optics, and $\unit[1.78]{\sigma}$ in flat optics. The latter being above the $\unit[1.5]{\sigma}$ limit, is a cause for concern. The beams would due to losses be extracted within three turns, and as explained in Fig.~\ref{fig:orbitBehavior}, the amplitude of the orbit perturbation would oscillate back and forth, spreading the losses over a longer time. Also, there is roughly a three-turn spread of the losses due to the betatron motion (fractional tune approximately one third). 

Nevertheless, to ensure the integrity of the machine, a detection of spurious QH discharges is required for the triplet magnets (Q1, Q2 and Q3), as well as the D1 separation dipoles. Furthermore, a detailed loss evaluation is necessary for Q2 and D1, and depending on the outcome of it, it might be required to ensure that the phase advance between these magnets and the primary collimators is kept away from $\unit[90]{\degree}$, in order to spread the losses over a longer time while still keeping the protection of the aperture.

These studies should be complemented by the beam screen shielding effect once it is understood, as it would provide a certain, unknown, delay to the onset of the losses.

\subsection{CLIQ}
CLIQ is only to be used in the three final focusing triplet magnets positioned near ATLAS (IP1) and CMS (IP5), designated Q1, Q2 and Q3, with Q1 being closest to the IP. As discussed above, the quench detection system will always ensure to trigger a beam dump before firing the magnet protection system. However, as for the QH, a spurious discharge of single CLIQ units cannot be excluded.

\begin{figure}[!htb]
  \begin{subfigure}{0.48\textwidth}
    \centering\includegraphics[width=0.92\textwidth]{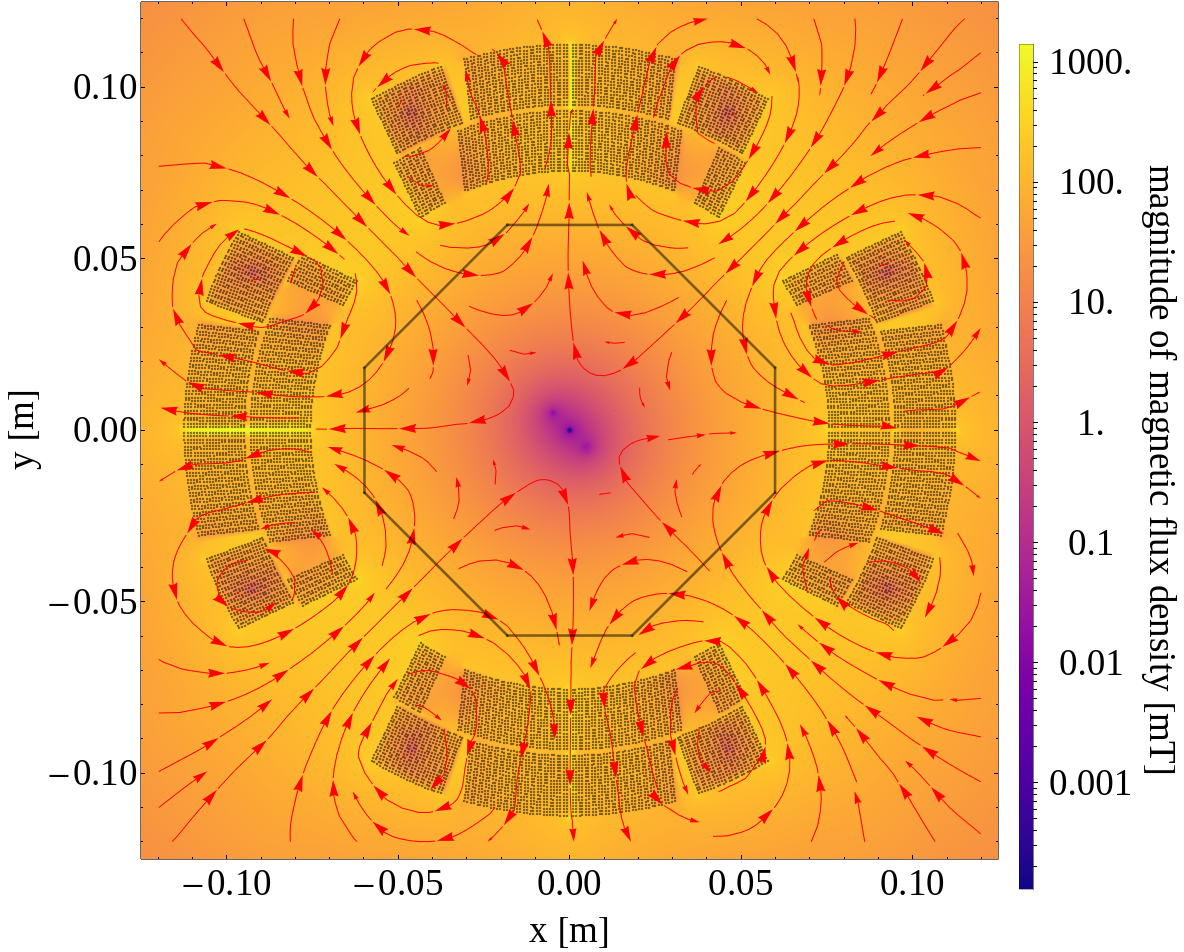}
   \caption{Q2 - skew octupolar field}
  \end{subfigure}
  \begin{subfigure}{0.48\textwidth}
    \centering\includegraphics[width=0.92\textwidth]{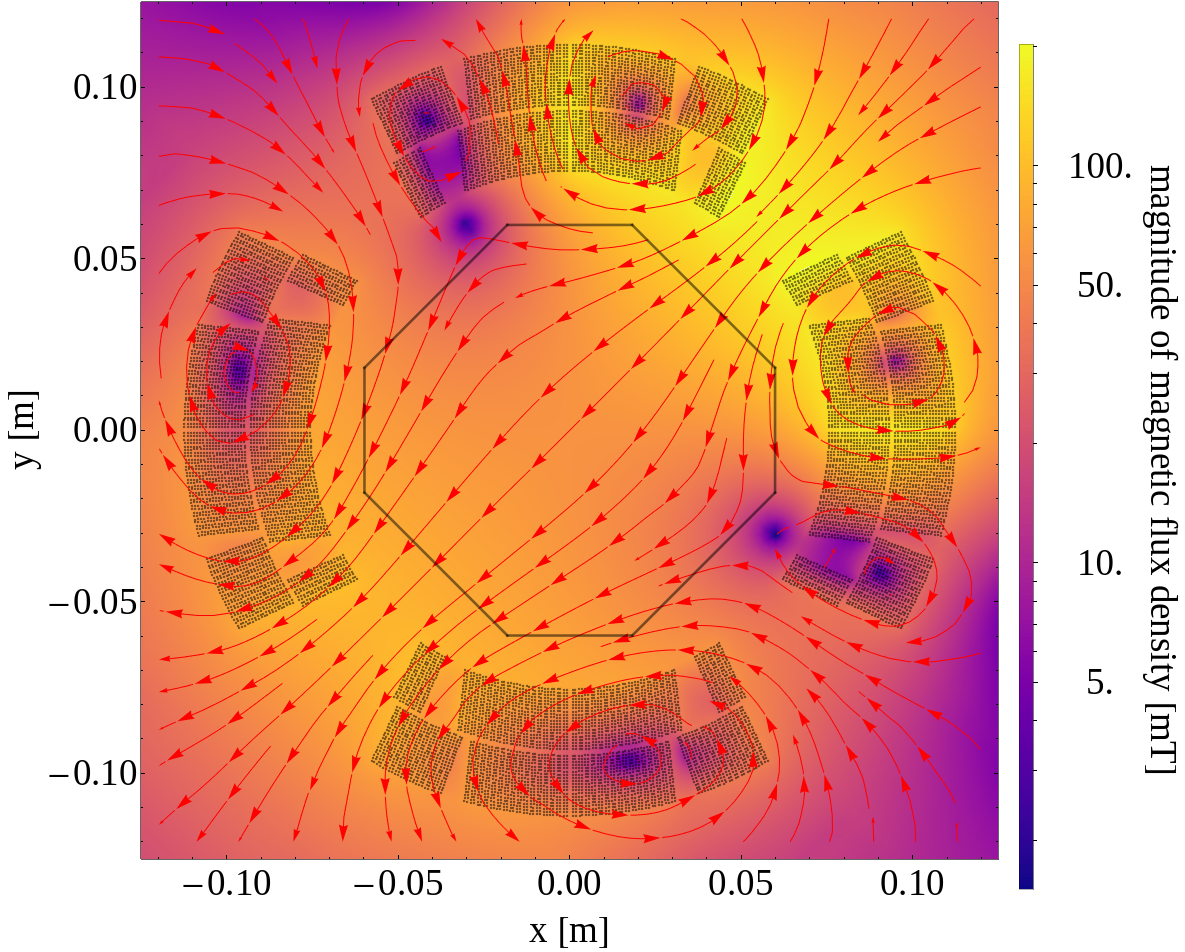}
   \caption{Q1/Q3 - normal and skew dipolar fields (the beam screen corresponds to Q3)}
  \end{subfigure}
  \caption{The magnetic field induced by a CLIQ discharge in the IP1 and IP5 final focusing triplet magnets, $\unit[5]{ms}$ after firing. The black lines show the beam screen and the black dots the cable positions. The CLIQ unit modulates the current in the main coil of the magnet.}
  \label{fig:CLIQmagFieldQ123}
\end{figure}

The currents in the magnets during a single CLIQ unit discharge, as well as the magnetic fields they induce, have been simulated using LEDET~\cite{LEDET} and SIGMA~\cite{SIGMA} of the STEAM~\cite{steam,bortot_steam_2018} framework. The interlock system is capable of detecting the CLIQ discharge within $\unit[500]{\mu s}$~\cite{denz_quench_2019}, putting a total delay from the firing of the CLIQ unit until the beams are dumped of up to $\unit[1]{ms}$. The peak of the CLIQ discharge occurs between $13$ and $\unit[25]{ms}$, and the main interest regarding its impact on the beam lies in the first few milliseconds. The resulting magnetic field is shown, as the difference between the field with CLIQ after $\unit[5]{ms}$ and the nominal magnetic field, in Fig.~\ref{fig:CLIQmagFieldQ123}. This cross-section is assumed to be constant along the length of the magnet and fringe fields are neglected due to their relatively small contribution.

The effect of the ramping CLIQ discharge on the beam was simulated using MAD-X~\cite{schmidt_mad-x_2006}, in order to take all optics changes into account as well as the dynamic behavior of the discharge itself. To apply the magnetic field in MAD-X, it was decomposed into normal and skew multipoles up to order 7 throughout the beam region $x,y\in[\unit[\pm25]{mm}]$. A summary of the multipole strengths for the three different magnets, $\unit[3]{ms}$ into the CLIQ discharge, is given in Table~\ref{tab:CLIQmultipole}. The bold values are those giving the largest kick on the beam. As can also be seen in Fig.~\ref{fig:CLIQmagFieldQ123}, for Q2, the main effect is from a skew octupolar field, whereas the kick from the normal quadrupolar field is less than $5~\%$ of this. In Q1 and Q3, the main components are the skew and normal dipolar fields, with skew and normal sextupolar fields giving a kick about $10~\%$ of the dipolar fields. The other components have a negligible effect on the orbit excursion and the beta beating.

\begin{table*}[t]\centering
  \caption{Summary of the multipolar decomposition $\unit[3]{ms}$ into the CLIQ discharge for all magnets without considering the shielding effect of the beam screen. The unit is in $\unit[]{T/m^{(n-1)}}$, where n designates the pole order. Q1 and Q3 are identical. The components giving the largest kick on the beam, for beam orbits up to $\unit[20]{mm}$, are in bold.}
  \begin{tabular}{lccccccccccc}
    \toprule
    \textbf{Multipole order} & 1 & 2 & 3 & 4 \\ \hline
    Q2 \\ 
    - normal components & $-3.36\times10^{-7}$ & $\mathbf{9.97\times10^{-3}}$ & $-4.13\times10^{-3}$ & $1.35$ \\%& $-76.9$ & $3.53\times10^3$ & $2.15\times10^6$ \\
    - skew components & $-1.28\times10^{-7}$ & $1.24\times10^{-5}$ & $-4.47\times10^{-3}$ & $\mathbf{-3.18\times10^3}$ %\\%& $15.0$ & $-5.69\times10^3$ & $1.87\times10^6$
    \\ \hline
    Q1/Q3 \\ 
    - normal components & $\mathbf{-2.1\times10^{-2}}$ & $-3.59\times10^{-5}$ & $\mathbf{-7.38}$ & $-5.37\times10^{-2}$ \\%& $7.53\times10^3$ & $1.67\times10^4$ & $2.82\times10^7$ \\
    - skew components & $\mathbf{-2.12\times10^{-2}}$ & $-2.66\times10^{-5}$ & $\mathbf{7.38}$ & $-1.68\times10^2$ %\\%& $7.57\times10^3$ & $7.86\times10^3$ & $-2.33\times10^7$
    \\ \hline
  \end{tabular}
  \label{tab:CLIQmultipole}
\end{table*}

\subsubsection{Results -- Round optics} 
\begin{figure}[!htb]
   \centering
   \includegraphics*[width=0.48\textwidth]{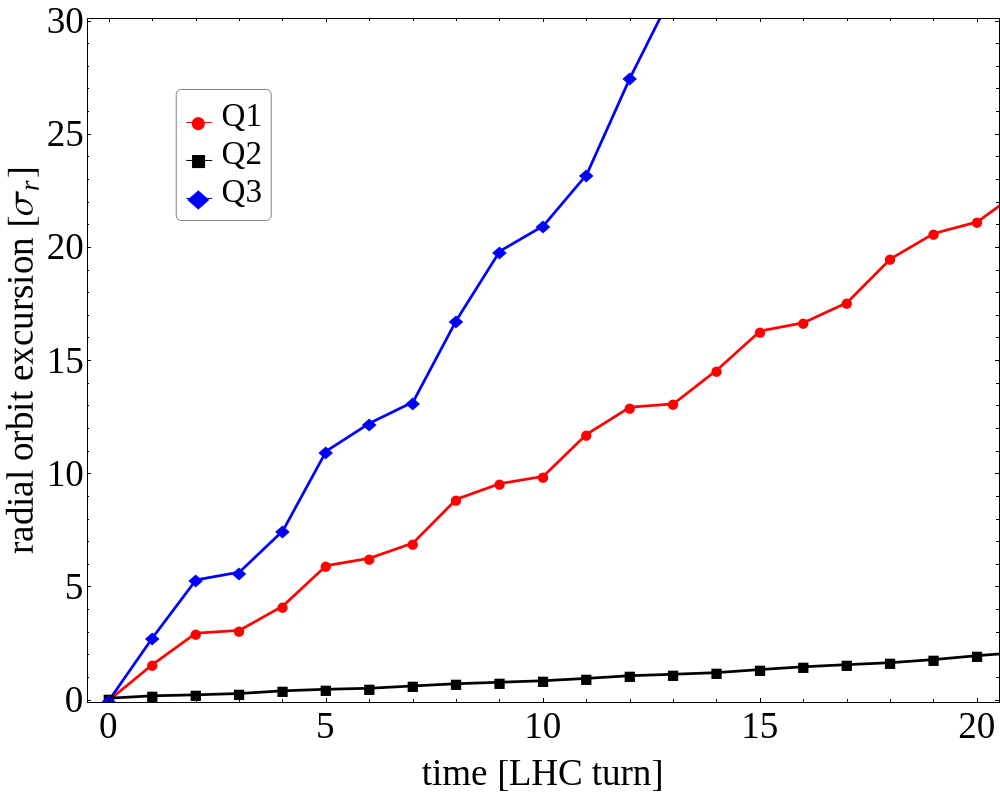}
   \caption{The orbit excursion induced by a CLIQ discharge in the three types of triplet magnets Q1, Q2 and Q3. For each type, only the magnet with the largest kick is shown.}
   \label{fig:CLIQoffset}
\end{figure}

In Fig.~\ref{fig:CLIQoffset}, the orbit excursion during the first 20 turns following a spurious discharge of a single CLIQ unit is shown for each of the three magnet types. As the kick depends on both the beta functions and on the beam orbit, due to the higher order components, the worst case of each type of triplet magnet was chosen for the plot. The horizontal and vertical orbit excursions were added in quadrature, to give the \textit{radial} orbit excursion. 

The critical orbit excursion of $\unit[1.5]{\sigma}$ is reached already in one turn for Q1 and Q3. There are no means of actively protecting the machine if such a spurious discharge were to happen and passive protection would not be sufficient, making this the worst known beam failure in the LHC.

For Q2, the effect on the beam is significantly slower, albeit still critical. The main reason for this is that the CLIQ unit is connected symmetrically around the magnet poles, such that no dipolar fields are produced. Furthermore, since the two magnet halves in Q2 are electrically split, only half the length of the full magnet is affected by the CLIQ discharge. Since the Q2 magnet halves are $\unit[7.15]{m}$ each and the Q1/Q3 magnet halves are $\unit[4.2]{m}$ each, this gives a small reduction as well.

To understand the difference in kick between the different magnets, the relevant optics parameters are listed in Appendix, Table~\ref{tab:tripletOptics}. In Q1 and Q3, since the kick is mainly due to a dipolar field, the different magnets of each type give similar kicks as the beta functions are of similar magnitude. Since the higher order components are also of equal magnitude in both the normal and the skew components, there is no meaningful difference between the magnets due to these either. For Q2 the kick is mainly due to an octupolar field component. As shown in Eq.~\ref{eq:kickIntegration}, the kick is only proportional to the square root of the beta function, but linearly proportional to the magnetic field change. This implies that the magnets with the largest orbit excursion give the largest kicks since the octupolar field is proportional to the cube of the orbit excursion, making the Q2s left of IP1 and right of IP5 (as seen from the center of the LHC) the worst cases.

\subsubsection{Mitigation -- Round optics}

\begin{figure}[!htb]
   \centering
   \includegraphics*[width=0.48\textwidth]{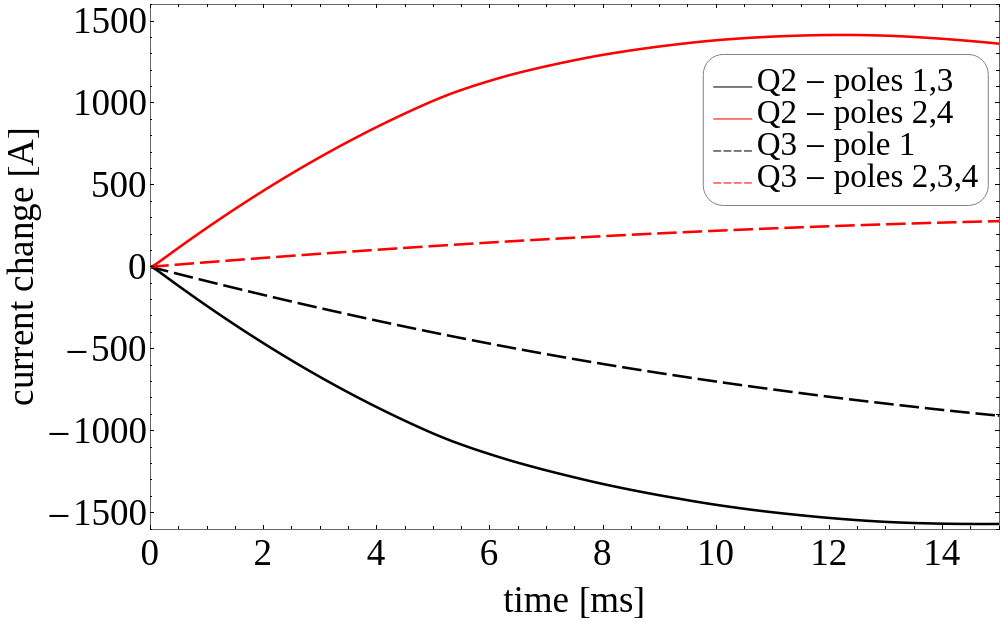}
   \caption{The current induced by CLIQ into the Q2 and Q3 magnets (Q1 is equivalent to Q3). In Q2, the change is symmetric between the two pairs of poles, whereas in Q3, one pole sees a much larger decrease of the current than the increase in the other three poles.}
   \label{fig:CLIQcurrent}
\end{figure}

\begin{figure*}[!htb]
  \begin{subfigure}{0.98\textwidth}
    \centering\includegraphics[width=0.92\textwidth]{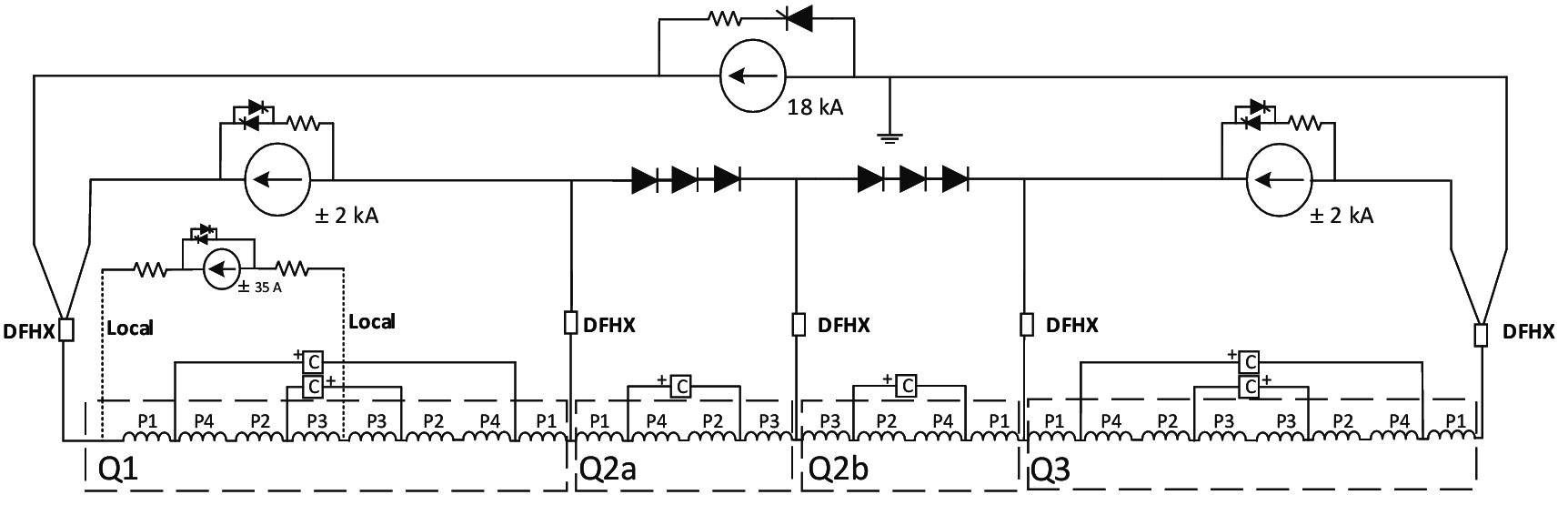}
   \caption{baseline in HLLHC TDR v. 0.1}
  \end{subfigure}
  \begin{subfigure}{0.98\textwidth}
    \centering\includegraphics[width=0.92\textwidth]{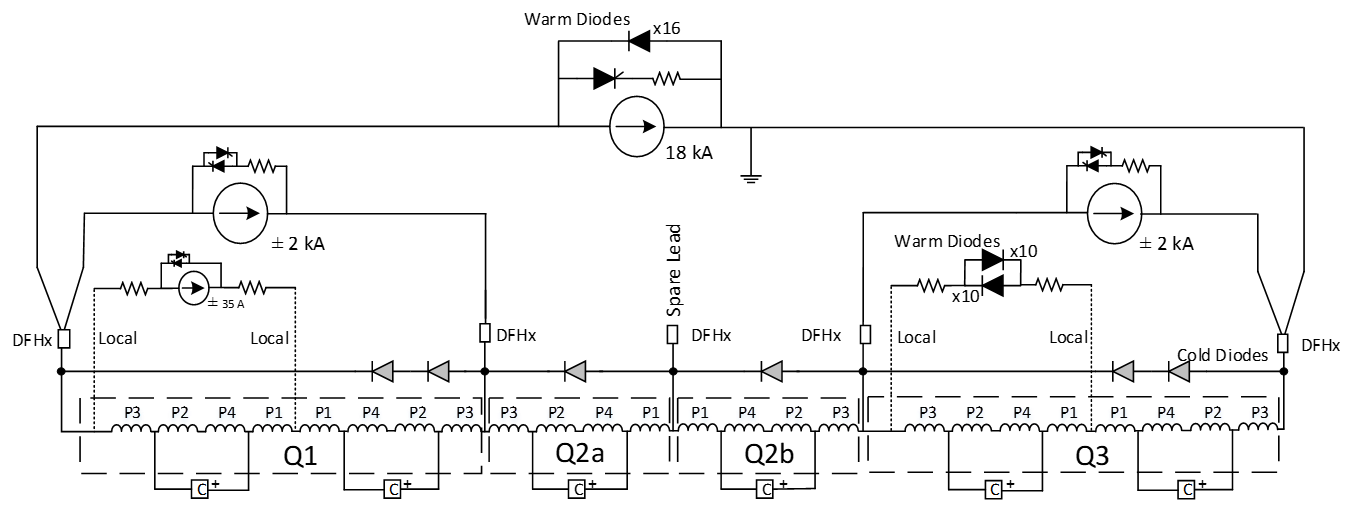}
   \caption{new baseline in HLLHC TDR v. 1.0}
  \end{subfigure}
   \caption{The baseline CLIQ connection scheme (top) as of the writing of this article~\cite{HLTDRv01_chapter7,MCF} vs the new baseline to be active from HLLHC TDR v. 1.0~\cite{HLTDRv10_chapter7}. The magnets, Q1, Q2 and Q3 are encircled by dashed lines. All three magnets consist of two identical halves, designated a and b, but only for Q2 the halves are separate. The magnet poles are designated by P1 to P4 and the CLIQ units by C.}
   \label{fig:CLIQcircuit}
\end{figure*}

The kick shown in the previous subsection for the Q1 and Q3 magnets is mainly an effect of the strong dipolar field components. The source of this is an asymmetry in the CLIQ discharge; as can be seen in Fig.~\ref{fig:CLIQcurrent} and the circuit in Fig.~\ref{fig:CLIQcircuit}~(a), the current change has the same polarity in three out of four poles in each magnet half when one CLIQ unit fires in Q1 and Q3. In Q2 the discharge is evenly distributed, with two poles seeing a reduction of the current and two poles seeing an equal increase of the current.

\begin{figure}[!htb]
   \centering
   \includegraphics*[width=0.48\textwidth]{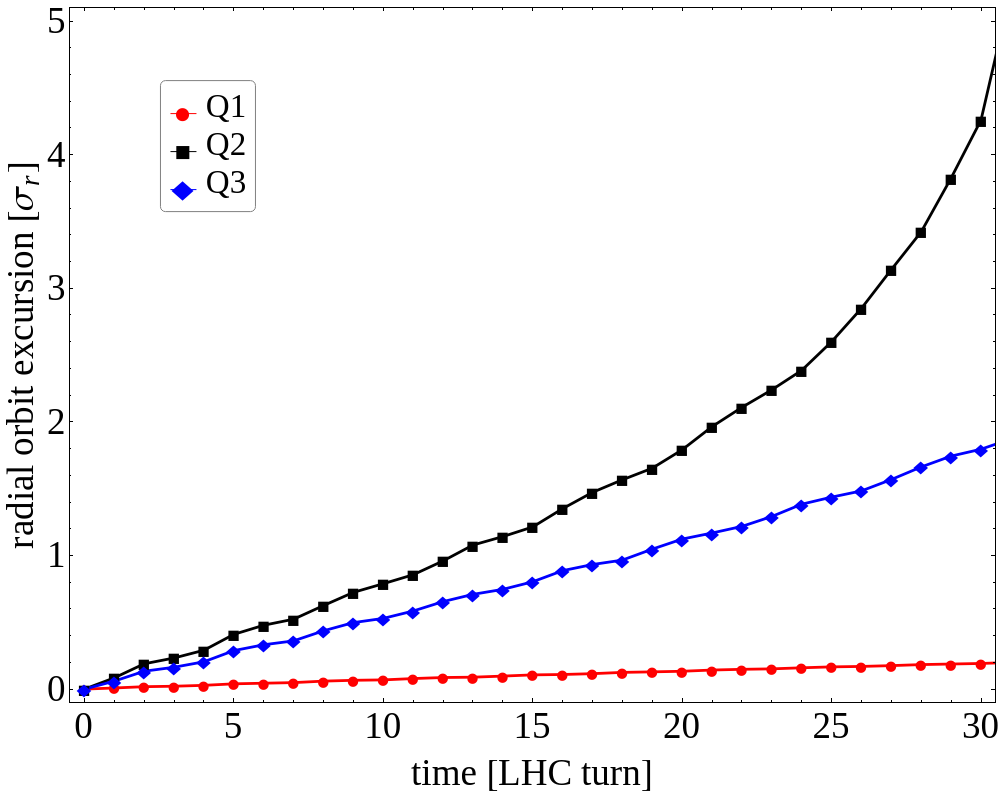}
   \caption{The orbit excursion induced by a CLIQ discharge using the new baseline in the three types of triplet magnets Q1, Q2 and Q3. For each type, only the magnet with the largest kick is shown.}
   \label{fig:CLIQoffsetNewSchematic}
\end{figure}

The kick in Q1 and Q3 can thus be significantly mitigated by changing the connection scheme of the CLIQ units to that of the Q2 magnets (Fig.~\ref{fig:CLIQcircuit}~(b)). This \textit{new} baseline schematic removes the strong dipolar field component, as well as decoupling the two magnet halves, halving the active length where CLIQ can act on the beam. The result is shown in Fig.~\ref{fig:CLIQoffsetNewSchematic}. Q2 would then constitute the worst case, reaching the $\unit[1.5]{\sigma}$ in 17 turns, and consequently requiring a fast and dedicated interlock system for spurious discharges.

\subsubsection{Beta beating -- Round optics}
Although the kick is reduced significantly by the new connection scheme, there is still strong beta beating induced by the discharge. Beta beating implies a number of detrimental consequences for the machine; the beam size in the collimators changes, inducing beam losses, and the hierarchy between different collimator types can be reversed. Since the collimation hierarchy is mainly a concern for longer-term cleaning, the main question is whether or not the beams can be dumped safely. It is also important that the direct losses onto the primary collimators due to an increased beam size do not exceed the limit set to $\unit[1]{MJ}$ of deposited energy.

Since the beta beating comes from the higher order magnetic components together with the beam offset, the magnets inducing the largest beta beating are those with a large orbit excursion, meaning the Q1 left of IP1 and right of IP5, the Q2 left of IP1 and right of IP5, the Q3 right of IP1 and left of IP5. 

However, as for the collimator hierarchies and effective collimator gap settings, this also depends on the phase advance between the magnet and the specific collimator, meaning that no general rule for the worst case can be defined. Hereinafter, the worst case magnet for each particular effect on the optics, is shown, separated into the three magnet types Q1, Q2 and Q3.

The beta beating was calculated iteratively in MAD-X, taking the LHC as a one-turn transfer line. The Twiss parameters at the end of one turn were used as input for the next turn, while the CLIQ field was being ramped.

The movable dump absorber~\cite{HLTDRv01_chapter14} is an absorber downstream of the dump kickers designed to intercept beam losses resulting from an asynchronous beam dump. If the effective setting of the dump absorber becomes larger than that of the tertiary collimators in IP1 and IP5, there is a risk that loss levels in the triplet magnets become too large, which must be prevented. The required margin between the dump absorber and the tertiary collimators depends on the phase advance between the beam extraction kickers and the tertiary collimators~\cite{bruce_calculations_2015,RoderikCollPhaseAdv}, since the beam would only pass the collimators once before being extracted during an asynchronous beam dump. This also implies that the margin between the dump absorber and the tertiary collimators is a soft limit, i.e. if the limit is just reached there is no concern.

In the worst case, the horizontal phase advance change from the beam extraction kickers to the tertiary collimators in IP5 is $\unit[-12]{\degree}$ within ten turns, for a CLIQ discharge in the Q2 right of IP5. This constitutes a significant change of margin of around $\unit[1]{\sigma}$.

\begin{figure}[!htb]
   \centering
   \includegraphics*[width=0.48\textwidth]{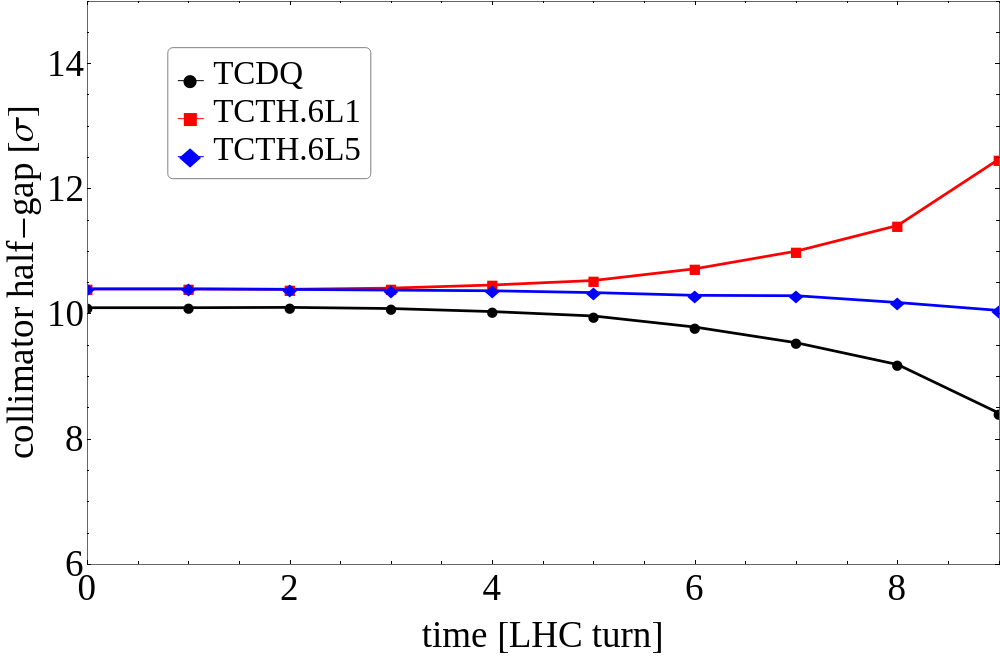}
   \caption{The effective collimator gap settings of the dump absorber (TCDQ) and the horizontal tertiary collimators (TCTH) in IPs 1 and 5 following a CLIQ discharge in the Q2 magnet left of IP1, constituting the fastest case - \textbf{round optics}.}
   \label{fig:CLIQ_TCDQmargin}
\end{figure}

In Fig.~\ref{fig:CLIQ_TCDQmargin} the change of the dump absorber and tertiary collimator settings solely due to the beta beating is shown for the fastest case, a CLIQ discharge in the Q3 left of IP1. The behavior is similar in all magnets, that is, that the effective setting of the dump absorber decreases while the effective setting of the tertiary collimators increase. This acts to increase the margin, offsetting the decrease due to the change of phase advance, and the protection consequently remains for all the cases.

\begin{figure}[!htb]
   \centering
   \includegraphics*[width=0.48\textwidth]{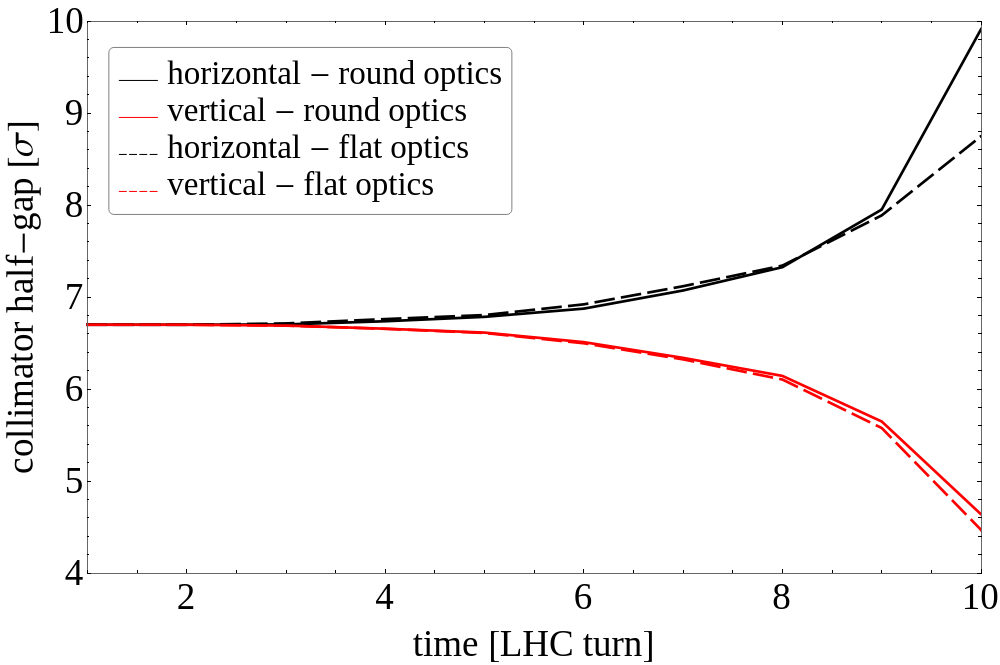}
   \caption{The effective gap settings of the horizontal and vertical primary collimators during a CLIQ discharge in the Q2 right of IP5, round optics, or the Q2 right of IP1, flat optics.}
   \label{fig:CLIQ_TCPmargin}
\end{figure}

As for the primary collimators, the largest decrease of the collimator gap occurs for CLIQ discharges in the Q2 right of IP5, and is shown in Fig.~\ref{fig:CLIQ_TCPmargin}. The change is large, with a reduction of the vertical gap by as much as $\unit[1.1]{\sigma}$ and $\unit[2.1]{\sigma}$ nine and ten turns respectively after the discharge starts.

The beams are offset vertically within this magnet, meaning that the net magnetic field vector will give a horizontal kick to the beam (c.f. Fig.~\ref{fig:CLIQmagFieldQ123}). Meanwhile, only the vertical collimator gap is reduced during the first ten turns. Since they are in different planes, the reduced collimator gap should not act to enhance the beam losses due to the orbit excursion. Furthermore, the direct losses due to the collimator gap reduction are cleaned out over more than three turns due to the betatron motion (tune is approximately one third). The beams will be dumped within ten turns, and the beta beating at the primary collimators is, in conclusion, not critical.

\subsubsection{Results -- Flat optics}
\begin{figure}[!htb]
   \centering
   \includegraphics*[width=0.48\textwidth]{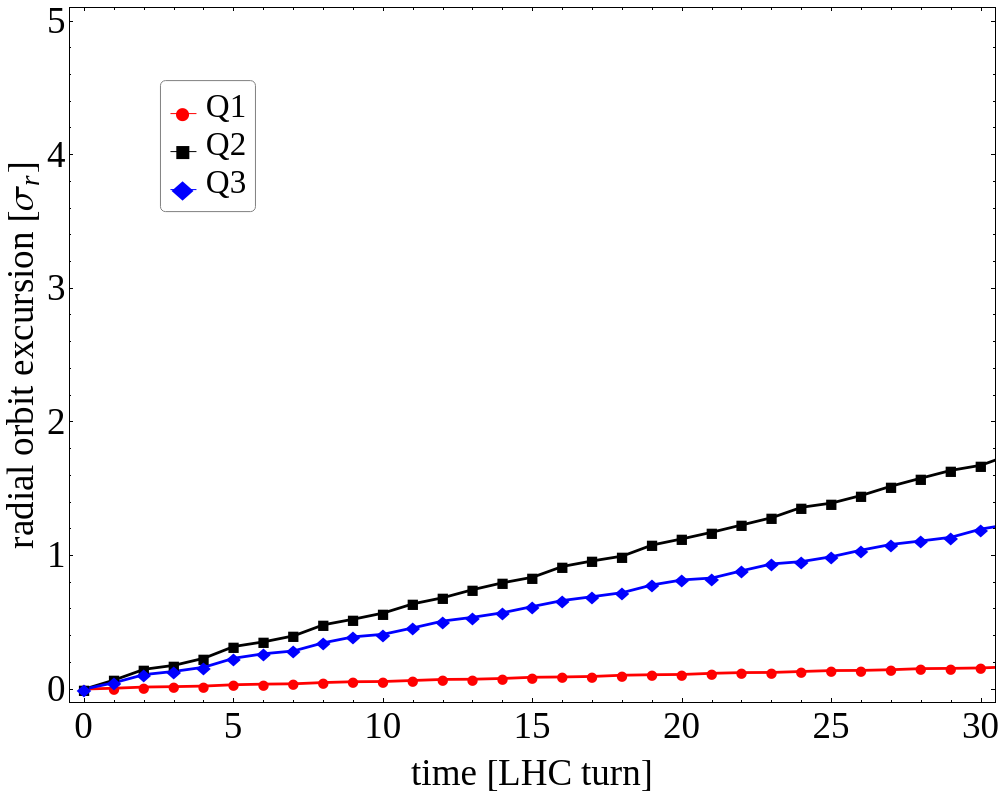}
   \caption{The orbit excursion induced by a CLIQ discharge using the new baseline in the three types of triplet magnets Q1, Q2 and Q3, using flat optics. For each type, only the magnet with the largest kick is shown. To be compared with the round optics case shown in Fig.~\ref{fig:CLIQoffsetNewSchematic}.}
   \label{fig:CLIQoffsetFlat}
\end{figure}

Figure~\ref{fig:CLIQoffsetFlat} shows a plot of the worst case orbit excursions for flat optics. The kick is mainly due to a skew octupolar field, meaning that the magnetic field seen by the beam is in the same direction as the offset due to the crossing angle, giving a kick perpendicular to this. Since the beta function is smaller in the crossing plane, but larger in the other plane, this increases the impact of a CLIQ discharge on the beam for flat optics vs the standard round optics. However, the beam orbits are also different, and in particular in the Q2 magnets, the orbit is a factor two smaller in flat optics, leading to roughly a factor eight smaller kick. The net effect is consequently that the impact of CLIQ in flat optics is smaller than in round optics. For Q2, the $\unit[1.5]{\sigma}$ limit is reached nine turns later with these optics, that is, after 26 turns.

Regarding beta beating, the results are in general similar to the round optics; in Fig.~\ref{fig:CLIQ_TCPmargin} the change of the primary collimator setting is compared for the two optics and the difference is negligible. 

\begin{figure}[!htb]
   \centering
   \includegraphics*[width=0.48\textwidth]{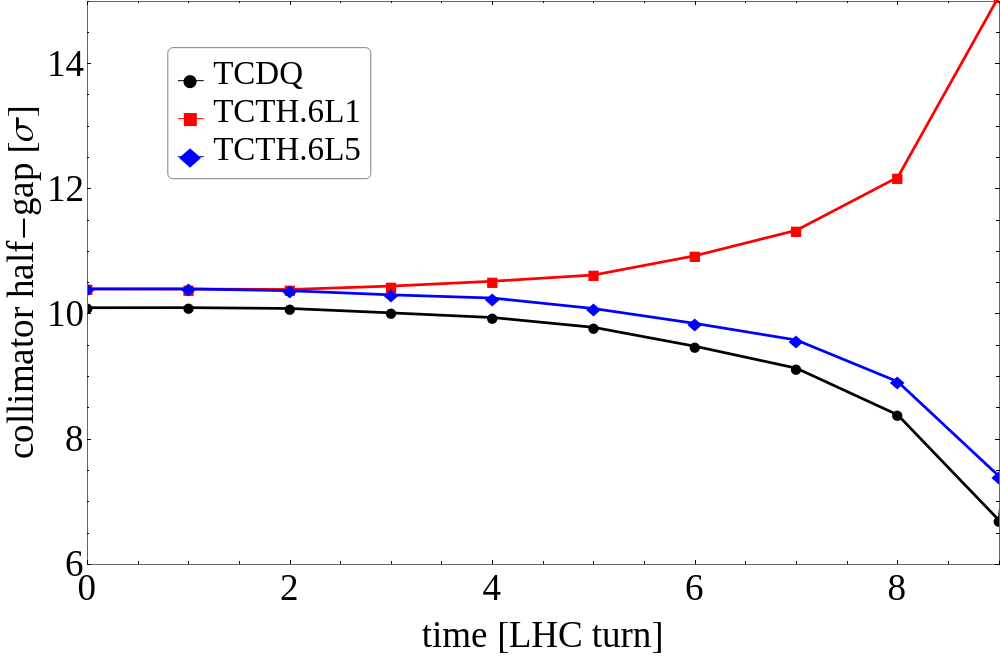}
   \caption{The effective collimator gap settings of the dump absorber (TCDQ) and the horizontal tertiary collimators (TCTH) in IPs 1 and 5 following a CLIQ discharge in the Q2 magnet left of IP1, constituting the fastest case - \textbf{flat optics}.}
   \label{fig:CLIQ_TCDQmarginFlat}
\end{figure}

For the margin between the tertiary collimator and the dump absorber, the behavior is similar to the round optics, in the sense that the dump absorber gap is reduced for all magnets, except for a discharge in the Q2 and Q3 of IP1, where the tertiary collimator margin in IP5 is decreased. The result for a discharge in the Q2 magnet right of IP1 is shown in Fig.~\ref{fig:CLIQ_TCDQmarginFlat}, where the margin is kept at around $\unit[0.5]{\sigma}$.

The change of phase advance between the beam extraction kickers and the tertiary collimator is $\unit[10]{\degree}$, for a discharge in the Q2 left of IP5 or Q2 right of IP1. For the tertiary collimator to the dump absorber margin, this implies a decrease of about $\unit[1]{\sigma}$~\cite{RoderikCollPhaseAdv}. Under this condition the margin between the tertiary collimators and the dump absorber is expected to be breached. Therefore, it needs to be carefully considered to increase this margin in case flat optics is used.

\subsubsection{Sensitivity on the beam orbit -- Round optics}
In the LHC, the orbit of the beam is interlocked by software to the level of $\unit[1]{mm}$~\cite{kain_injection_nodate} with respect to the reference orbit, and there are fill-by-fill variations. Furthermore, the reference orbit will be offset in relation to the design orbit of the optics~\cite{2mmAssumption}. It is thus assumed that the beams could be up to $\unit[2]{mm}$ off under normal conditions in both transverse directions separately, inside the triplet magnets. Since the field due to the CLIQ discharge consists of higher-order components, the beam orbit plays a significant role; with a larger orbit excursion, the field, and consequently the kick, would be enhanced. This is also important to consider for any changes in optics, e.g. an increase of the crossing angle, that would lead to larger orbits in the triplet magnets.

\begin{figure}[!htb]
   \centering
   \includegraphics*[width=0.48\textwidth]{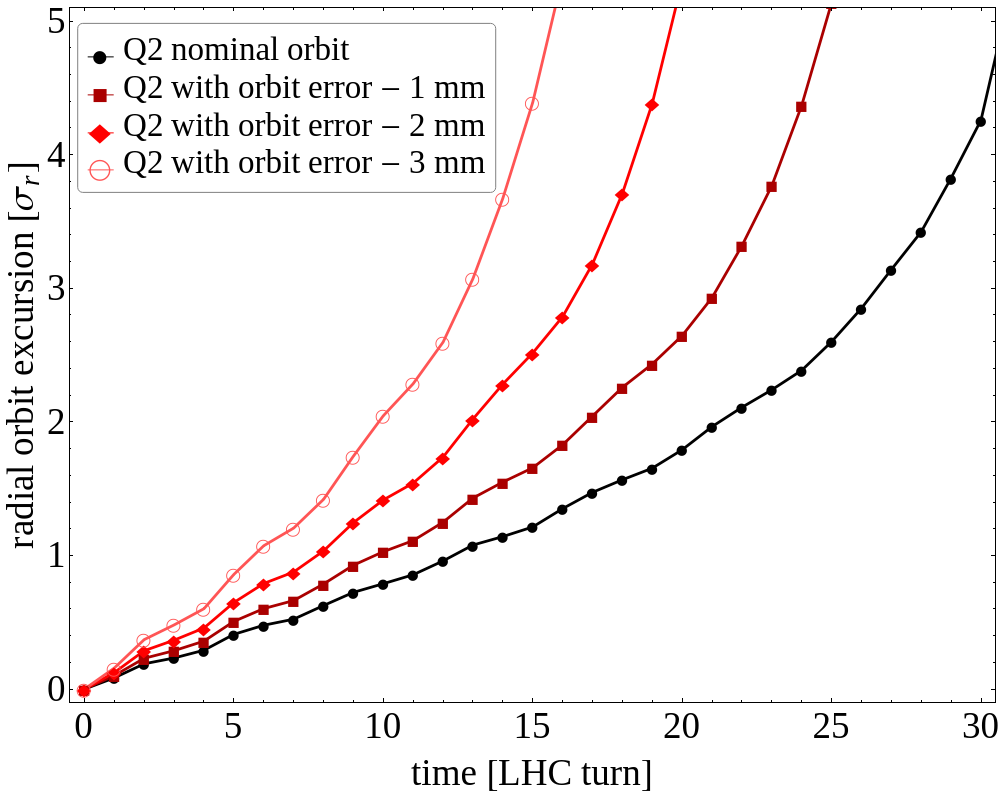}
   \caption{Comparison of different orbit errors and their effect on the orbit excursion resulting from a CLIQ discharge in the worst magnet. Only round optics is considered, and the orbit errors are applied both horizontally and vertically simultaneously.}
   \label{fig:CLIQ_offsetWithError}
\end{figure}

The worst case Q2 magnet was simulated assuming that the beam orbit was further displaced from the center of the magnet in both transverse directions. As seen in Fig.~\ref{fig:CLIQ_offsetWithError}, this increases the orbit excursion, and the $\unit[1.5]{\sigma}$ limit is reached on turn 14 instead of turn 17 for an initial offset of $\unit[1]{mm}$. This is still longer than the time required to dump the beams. If the initial offset instead is increased to $\unit[2]{mm}$, the limit is reached already after 10 turns, leaving just enough time to dump the beams. An offset of $\unit[3]{mm}$ makes the event too fast for the expected interlock time and needs to be avoided.

\subsubsection{CLIQ Conclusions}
CLIQ is necessary for effective protection of the HL-LHC era final focusing triplet magnets against damage induced by quenches. However, due to the large currents of up to $\unit[1.5]{kA}$ induced directly into the magnet coils, the effects on the beam can be significant in case of spurious triggering. Some of these cases constitute the worst known possible failures in the LHC, and as for the QHs, spurious discharges of single CLIQ units cannot be fully excluded. It is thus required that the CLIQ units are interlocked against spurious discharges, that their impact on the beams in case of spurious discharges is minimized, and that the quench protection system ensures complete extraction of both beams before CLIQ is triggered.

Under the current baseline, the CLIQ units in the Q1 and Q3 triplet magnets produce strong dipolar fields, giving an orbit excursion of almost $\unit[3]{\sigma}$ already after the first turn, meaning that the beam core would impinge onto the primary collimators if the phase advance is close to $\unit[90]{\degree}$. After three turns, the dump time due to a trigger by the fastest beam loss monitors, the orbit excursion would already be at $\unit[6]{\sigma}$. 
There are no means of protecting against this kind of failure. Therefore the circuit of the CLIQ unit connection to the Q1 and Q3 magnets has been modified to that of the Q2 magnet. This removes the dipolar field components, and the orbit excursion instead comes from a skew octupolar field, resulting in a fast, yet tolerable, orbit excursion of $\unit[1.5]{\sigma}$ in 17 turns. This mitigation is being adopted as the new baseline~\cite{HLTDRv10_chapter7}.

For flat optics, these kicks are limited further, with $\unit[1.5]{\sigma}$ being reached in 26 turns. 

As for beta beating, the primary collimator gap settings are only reduced vertically for both round and flat optics, down to about $\unit[6]{\sigma}$, nine turns after the onset of CLIQ. With a detection and dump delay of less than ten turns, this is not a concern. The tertiary collimator to the dump absorber margin is kept for round optics. However, for flat optics this margin is not sufficient and it needs to be carefully considered to increase it. 

In case the beams have a larger nominal orbit inside the triplets, the CLIQ effect is enhanced due to the octupolar field being the dominating component. In particular, at an orbit offset of $\unit[2]{mm}$, horizontally and vertically, the limit for orbit excursion is reached already in ten turns. It is thus required that the optics are not changed such that a beam of unsafe beam intensity has a larger than nominal offset in the triplet magnets. The real triplet orbit should be measured in detail and interlock limits set such that it cannot deviate more than $\unit[2]{mm}$ from the design orbit in either plane for the baseline round optics, or $\unit[1]{mm}$ for flat optics. Reduced orbits (beam separations) within the triplet magnets are however not a concern.

\section{Coherent Beam-Beam Kick}
\label{BBK}
The coherent Beam-Beam Kick (BBK) is an electromagnetic interaction between counter-circulating bunches, which results in a transverse kick on the beam. This kick is expected and compensated for by correction magnets, leaving an effective overcorrection when only one beam is dumped. In the HL-LHC, the loss of the BBK due to the dump of one beam can consequently move the other beam by more than $\unit[1]{\sigma}$~\cite{tobias}. In the LHC there are separate \textit{beam permits} for the two beams, allowing them to be injected and circulate. During high intensity beam operation these beam permits are \textit{linked}, i.e.\ if a fault requiring a protection dump of one beam occurs, the second beam is dumped as well. Nevertheless, the dump of the second beam can be delayed by up to three turns~\cite{StephaneMPP}. 

\begin{figure}[!htb]
   \centering
   \includegraphics*[width=0.48\textwidth]{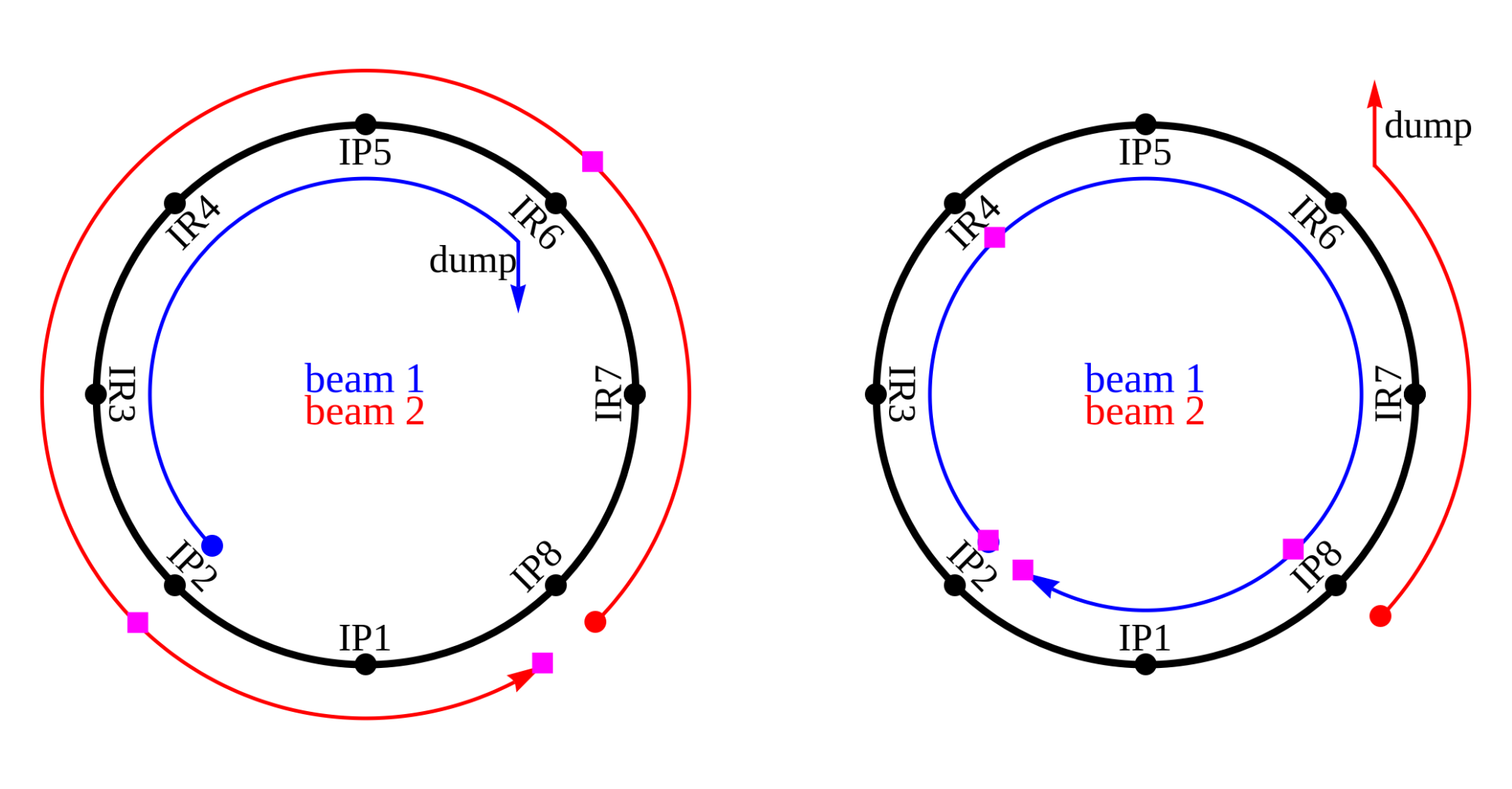}
   \caption{Schematic top view of the dump of the two beams; for simplicity Beam~1 is shown as the inner and Beam~2 as the outer one. The arrow indicates the beam direction, and the magenta squares show how the beams are split into three different parts when it comes to the BBK. The depicted abort gap is $\unit[3]{\mu s}$ long, and one turn is $\unit[89]{\mu s}$.}
   \label{fig:BBKdumpSchema}
\end{figure}

In Fig.~\ref{fig:BBKdumpSchema} a schematic view of the two beams during the dump of the other beam is shown. The dump occurs in IR6, whereas the loss of BBK occurs in IPs 1, 2, 5 and 8 as this is where both beams share a common beam pipe. Since the abort gaps are only synchronized in IPs 1 and 5, both beams cannot be dumped at exactly the same time and a loss of the BBK is experienced on at least one turn for parts of one of the beams. As can be deduced from the illustration, the two beams start seeing a loss of the BBK in different IPs. This is also true for three different parts of each of the beams; the first quarter, the middle half, and finally the fourth quarter (indicated by the magenta squares in Fig.~\ref{fig:BBKdumpSchema}). In which IPs the different parts of the beam is observed to have lost the BBK, on a given turn, depends on where one is observing the beam.

The resulting orbit excursion leads to beam losses in the betatron collimation region (IR7). These losses start already on the same turn as the dump of the first beam starts; from the start of Beam~1, or from the last quarter of Beam~2, depending on which beam is dumped first. If the other beam is dumped as soon as its abort gap reaches IR6, no losses would be seen from Beam~1, whereas the full Beam~2 would give losses on one turn. The last quarter of Beam~2 would give losses on two turns. 

In IR4, where the beam position monitors (BPMs) used for the measurements of the BBK analyzed in this section, are located, no effect is seen on Beam~1 and only a small kick from IP8 on the last quarter of Beam~2 is seen, if the remaining beam is dumped as soon as possible. For longer delays, Beam~2 would be observed in IR4 as follows:
\begin{itemize}
    \item \textbf{Beam part 1} - On turn 1: no loss of BBK. On turn 2: loss of BBK in IPs 1, 8 and 5. On Turn 3: loss of BBK in IP2.
    \item \textbf{Beam part 2} - On turn 1: no loss of BBK. On turn 2: loss of BBK in IPs 2, 1, 8 and 5. On turn 3: no loss of BBK.
    \item \textbf{Beam part 3} - On turn 1: loss of BBK in IP8. On turn 2: loss of BBK in IPs 2 and 1. On turn 3: loss of BBK in IP5.
\end{itemize}

\begin{figure}[!htb]
   \centering
   \includegraphics*[width=0.48\textwidth]{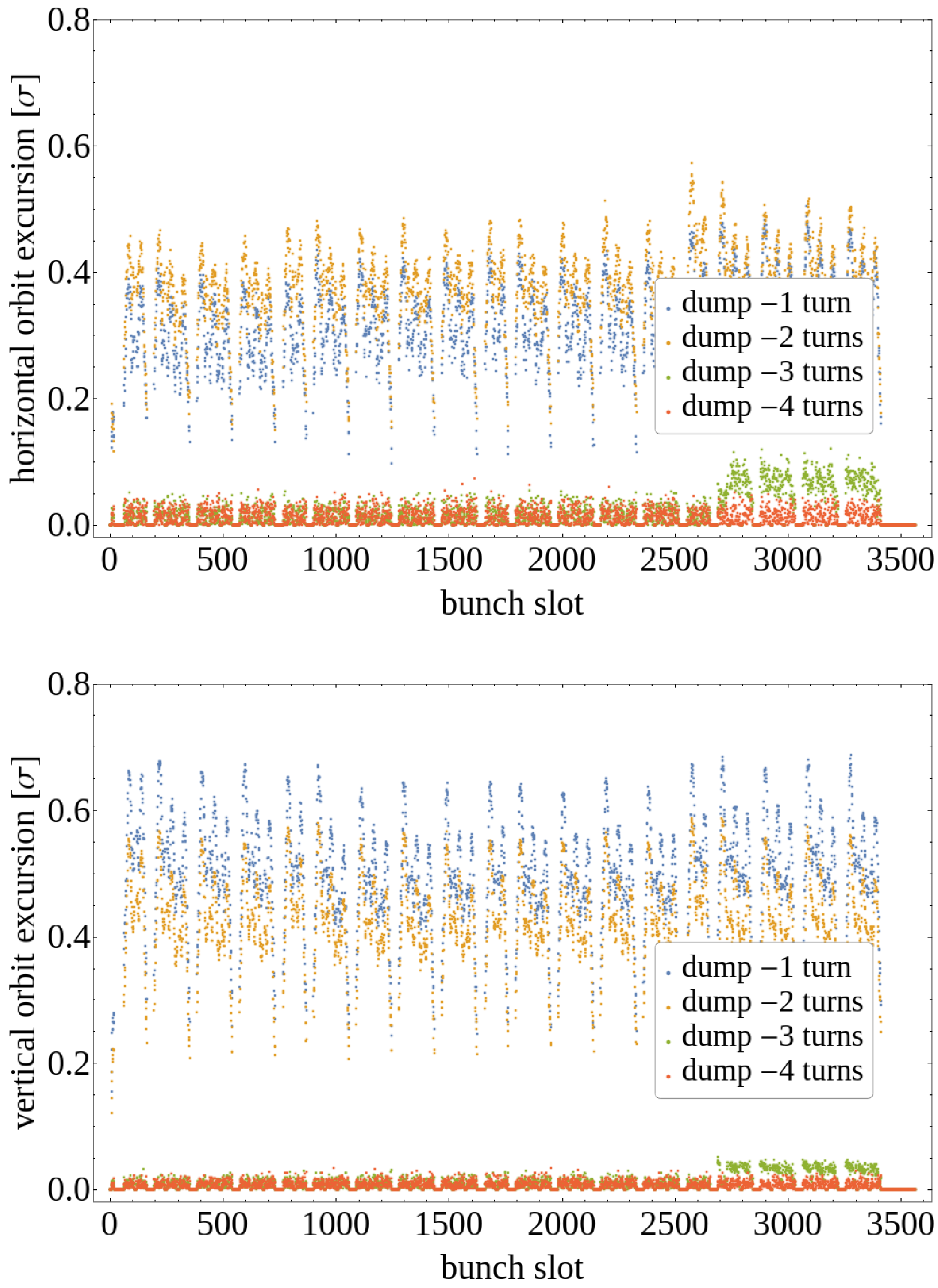}
   \caption{Measured horizontal (top) and vertical (bottom) orbit excursion of Beam~2 during the last four turns before beam dump of fill 7217. The effect of the loss of the BBK is clearly seen, with a vertical offset reaching $\unit[0.6]{\sigma}$ and a horizontal offset reaching $\unit[0.5]{\sigma}$.}
   \label{fig:BBKf7217}
\end{figure}

After analyzing all fills in 2017 and 2018, a total of 36 events during ramp or at top energy, where a loss of the BBK was present for two turns (28 events) or three turns (8 events), were found. An example can be seen in Fig.~\ref{fig:BBKf7217}, showing the normalized orbit excursion in units of $\sigma$ for the last four turns of Beam~2. The red line (dump minus 4 turns) shows the baseline, whereas the green line corresponds to the turn when Beam~1 starts being dumped. There is a clear effect only on the last quarter of Beam~2 on this turn, coming from a loss in IP8. IP8 has a diagonal crossing, but it is mainly horizontal, which is why the kick is strongest in the horizontal plane. On the following two turns, before Beam~2 is also dumped, the orbit excursion is seen for the full beam, reaching around $\unit[0.6]{\sigma}$. While this is not critical in the current LHC, given the larger bunch intensity and different optics in the future, it is important to evaluate its criticality in the LHC Run III and HL-LHC eras.

\subsection{Simulations}
In order to evaluate the magnitude of the BBK in the LHC during Run III and HL-LHC, simulations were done using the \textit{beambeam} module in MAD-X. As has been done previously~\cite{tobias}, the loss of the kick was simulated by tracking a few turns without BBK and then applying a negative BBK. This was reiterated with the correct timing of the loss in the different IPs, for the three different parts of each beam, and the latest version of the HL-LHC optics. A single reference particle with the nominal parameters is tracked in all cases, and the filling pattern is ignored, implying that the particle sees the BBK from all buckets every $\unit[25]{ns}$. The simulations were benchmarked against the measurements using LHC Run II optics and beam parameters. For the benchmarking, the tracked particle was observed at the location of the BPMs in IR4, but for the evaluations of the criticality in the future machine settings, it was observed at the primary collimators. The coherent BBK in the bunch center is zero, and approximately linear in the vicinity, consequently having a negligible effect in the collision point. This \textit{head-on} BBK was thus neglected to avoid numerical instabilities.

\begin{figure}[htb!]
  \begin{subfigure}{0.48\textwidth}
    \centering\includegraphics[width=0.98\textwidth]{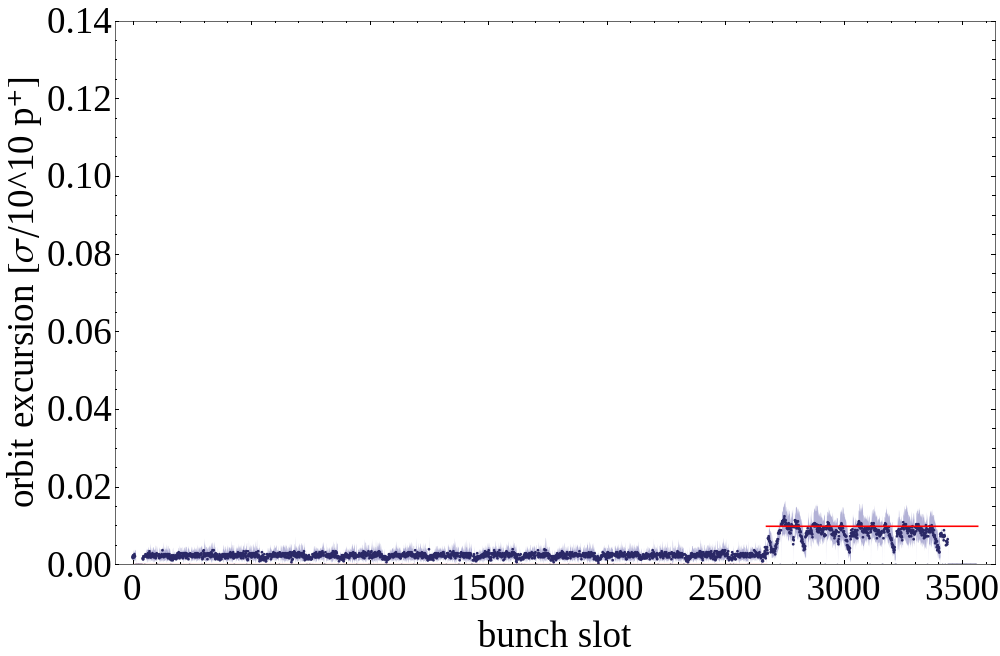}
    \caption{Three turns before dump}
  \end{subfigure}
  \begin{subfigure}{0.48\textwidth}
    \centering\includegraphics[width=0.98\textwidth]{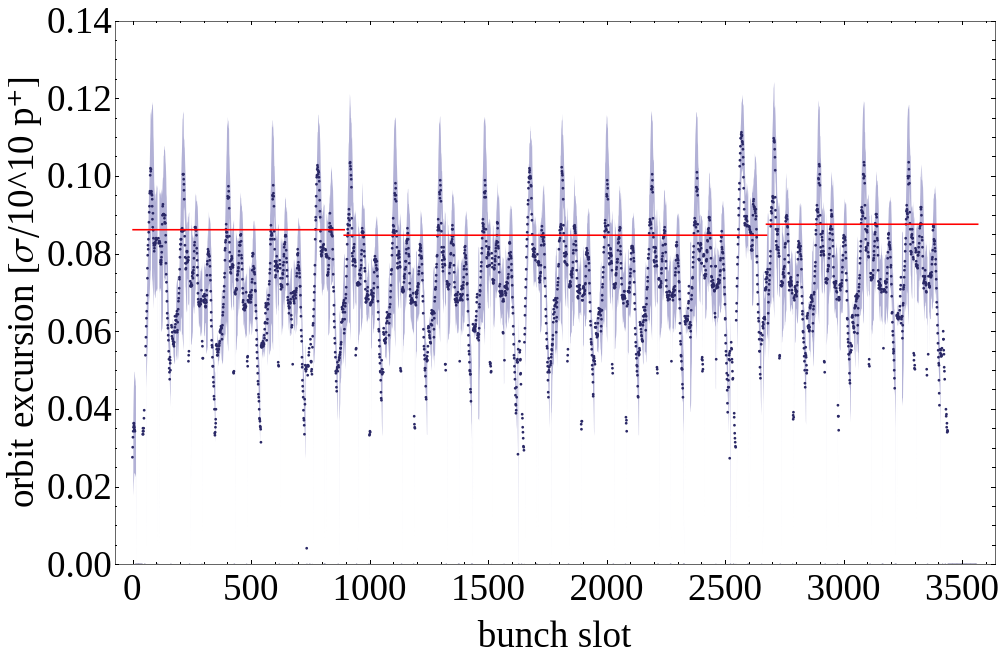}
    \caption{Two turns before dump}
  \end{subfigure}
  \begin{subfigure}{0.48\textwidth}
    \centering\includegraphics[width=0.98\textwidth]{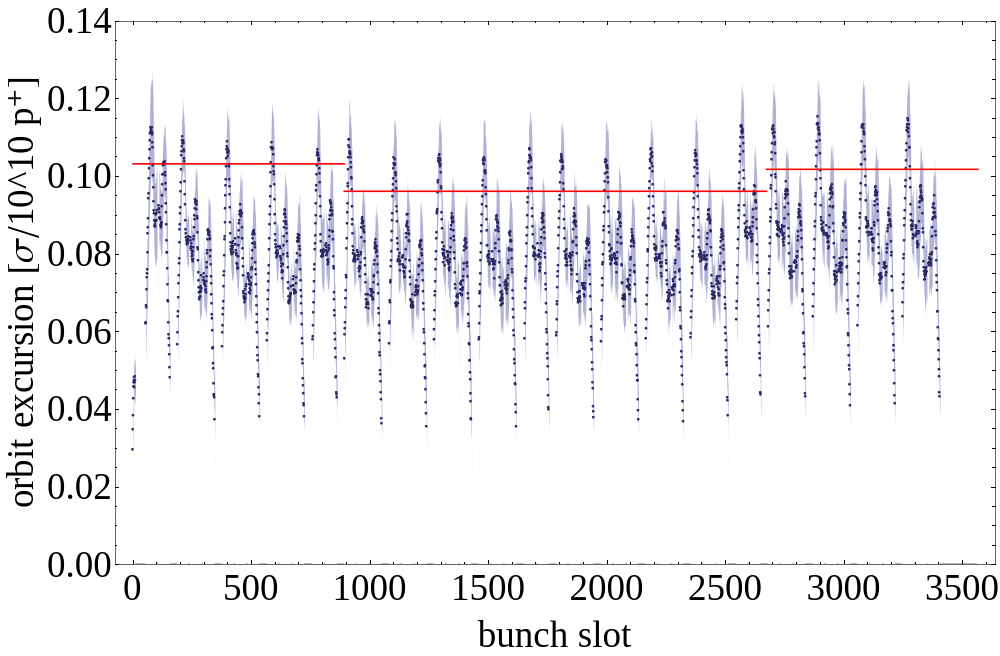}
    \caption{One turn before dump}
  \end{subfigure}
  \caption{Comparison between the measurements and the simulations of the loss of the BBK in Beam~2. The blue dots show the mean of all the 14 measurements at $\unit[6.5]{TeV}$, with the shaded region around them indicating one standard deviation. The red lines show the estimates from the simulations of the three beam parts.} 
  \label{fig:BBKcomparison}
\end{figure}

A comparison of the simulated and measured kicks for all Beam~2 events is shown in Fig.~\ref{fig:BBKcomparison}, where the vertical axis shows the radial orbit excursion in units of \textsigma~normalized to the bunch intensity. The influence of the emittance variation was neglected, since it is small compared to the bunch intensity variation. 
In the simulations, the full BBK was assumed, but in reality, different bunches see a different number of BBKs due to their position in the filling scheme; some of the encounters might be with empty buckets instead of opposing bunches. This also means that the simulations provide a conservative estimate on the resulting orbit excursions, a more precise estimate would require simulating the unique conditions of every single bunch. The 12 bunch train in the beginning of the filling pattern has been treated like nominal bunches, but normally their parameters (intensity and emittance) differ from the other bunches in the beam. Also, the 12 bunch train does not collide in the IPs. This is why they have a lower amplitude in the measurements.

\subsection{Criticality in LHC Run III and HL-LHC}
The criticality of the loss of the BBK for future LHC operations was evaluated through the simulations presented here. A summary of the beam parameters used for the simulations for LHC Run III and HL-LHC can be seen in Table~\ref{tab:beamParams}. In Run III, the main differences to LHC Run II will be an increase of the bunch intensity from $1.15\times10^{11}$ to $1.8\times10^{11}$ protons per bunch, as well as a decrease of the crossing angle to $\unit[109]{\mu rad}$. In HL-LHC the bunch intensity will be further increased to $2.2\times10^{11}$ protons per bunch, but the crossing angle will be increased, limiting the strength of the BBK.

\begin{table*}[bt]
   \centering
   \caption{Comparison of the optics parameters used for Run III~\cite{runIIIoptics} and HL-LHC (\textit{HLLHCv1.4}~\cite{maria_high_2019}) simulations of the Beam-Beam Kick. The crossing orientations are Vertical (V) or Horizontal (H).}
    \begin{tabular}{llll|lll}
                                        & \multicolumn{3}{c}{Run III} & \multicolumn{2}{c}{HL-LHC} &  \\
                                        & Start SB      & End SB & Worst case & Round   & Flat       &  \\
                                        \toprule
        \textbeta\textsuperscript{*}~IP1~[cm]  & 105           & 30 & 30        & 15      & 7.5/30     &  \\
        \textbeta\textsuperscript{*}~IP2~[cm]  & 1000          & 1000   & 1000     & 1000    & 1000       &  \\       
        \textbeta\textsuperscript{*}~IP5~[cm]  & 105           & 30 & 30        & 15      & 7.5/30     &  \\
        \textbeta\textsuperscript{*}~IP8~[cm]  & 150           & 150    & 150       & 150     & 150        &  \\ 
        crossing~IP1~[\textmu rad] & 109 V       & 162 V        & 162 V    & 295 H & 245 V    &  \\
        crossing~IP2~[\textmu rad] & 200 V         & 200 V              & 200 V    & 270 V   & 270 V         &  \\
        crossing~IP5~[\textmu rad] & 109 H       & 162 H        & 162 H    & 295 V & 245 H    &  \\
        crossing~IP8~[\textmu rad] & 250 H         & 250 H              & 250 V    & 115 H   & 115 H         &  \\
        bunch~intensity~[$\unit[10^{11}]{p^+}$]    & 1.8  & 1.2         & 1.5      & 2.2     & 2.2        &  \\
        RMS~bunch~length~[ns]         & 0.3           & 0.25            & 0.3   & 0.25    & 0.25       &  \\
        \textepsilon\textsubscript{n}~[\textmu m rad]   & 2.5  & 2.5    & 2.5         & 2.5     & 2.5        &  \\
        energy~[TeV]                  & 6.5/7         & 6.5/7           & 6.5/7     & 7       & 7          &  \\
                                        \bottomrule
    \end{tabular}
    \label{tab:beamParams}
\end{table*}

For Run III, three scenarios were considered: 

\begin{itemize}
    \item Beam parameters and optics during the start of collision directly after top energy has been reached and the beams have been squeezed to their collision parameters (referred to as Stable Beams, SB).
    \item Beam parameters and optics at the end of SB which is the point when the beams are dumped and the cycle is finished.
    \item Lastly a realistic combination of the parameters of both cases constituting a worst case scenario is considered, where the fully squeezed optics are used with a larger than nominal bunch intensity.
\end{itemize}
For HL-LHC, levelling~\cite{levelling} has not been considered since the fully squeezed optics could be used with nominal bunch intensities and the machine should be designed to handle it safely.

The normalized kick resulting from each IP individually is shown in Table~\ref{tab:bbkResultsIPseparate}. Both beams see the same normalized kick, and the kick is only in the crossing plane, meaning horizontal for IP1, vertical for IP2, vertical for IP5 and horizontal for IP8. IPs 1 and 5 give the strongest kick due to the smaller \textbeta\textsuperscript{*}, since the BBK increases with smaller beam size. The small \textbeta\textsuperscript{*} also implies larger beta functions around the IP, increasing the normalized kicks. Lastly, for LHC Run III, the crossing angle is smaller in IPs 1 and 5, than in 2 and 8, which further increases the kick.

\begin{table}[!htb]
   \centering
   \caption{Results of the BBK simulations per IP for Run III and HL-LHC, shown as the \textbf{radial orbit excursion in units of beam \textsigma}. The values are the orbit excursion that the remaining beam would see on the first turn if it looses the BBK of one individual IP.}
    \begin{tabular}{lcccc}
        Optics  &   IP1 (hor.) & IP2 (ver.) & IP5 (ver.) & IP8 (hor.) \\
        \toprule
        \textbf{Run III}& & &  \\
        Start of SB & 0.79 & 0.18 & 0.43 & 0.23 \\
        End of SB   & 0.65 & 0.08 & 0.52 & 0.15 \\  
        Worst case  & 0.81 & 0.11 & 0.65 & 0.19 \\
        \midrule
        \textbf{HL-LHC}& & &  \\
        Round & 0.86 & 0.09 & 0.86 & 0.58 \\
        Flat & 0.72 & 0.09 & 0.72 & 0.57 \\
        \bottomrule
    \end{tabular}
    \label{tab:bbkResultsIPseparate}
\end{table}

\begin{table}[!htb]
   \centering
   \caption{BBK simulations results, combining the IPs, for Run III and HL-LHC shown as the \textbf{radial orbit excursion in units of beam \textsigma}. Run II optics were used and adapted to Run III parameters. For HL-LHC the \textit{HLLHCv1.4} optics were used. The weighted average of the orbit excursion for the three different beam parts, as defined above, is shown, over three turns.}
    \begin{tabular}{lccc|ccc}
        & \multicolumn{3}{c}{\textbf{Beam~1}} & \multicolumn{3}{c}{\textbf{Beam~2}} \\
        Optics  &   turn 1 & turn 2   & turn 3 &   turn 1 & turn 2   & turn 3 \\
        \toprule
        \textbf{Run III}& & &  \\
        Start of SB & 0.44 & 0.74 & 0.70 & 0.05 & 0.88 & \textbf{1.03} \\
        End of SB   & 0.52 & 0.89 & 0.76 & 0.04 & 0.69 & \textbf{1.00} \\  
        Worst case  & 0.65 & \textbf{1.12} & 0.94 & 0.05 & 0.86 & \textbf{1.24} \\
        \midrule
        \textbf{HL-LHC}& & &  \\
        Round & 0.86 & \textbf{1.60} & 1.56 & 0.19 & 1.27 & \textbf{1.48} \\
        Flat & 0.72 & 1.21 & 0.89 & 0.19 & 0.95 & 1.39 \\
        \bottomrule
    \end{tabular}
    \label{tab:bbkResults}
\end{table}

The results of the full simulations, considering a loss of the BBK in all IPs combined, adhering to their respective timings as explained above, are shown in Table~\ref{tab:bbkResults}. Due to the two beams seeing a loss of the BBK in a different order of the IPs, the amplitude of the orbit excursion is different, and the two beams are shown separately. Both beams are observed in IR7, where the betatron collimation region lies, and consequently where the main losses are observed. The amplitude of the orbit excursion seen here is the same in IR6, which is where the beams are extracted to the beam dump. This is because there is no shared beam pipe in or between IRs 6 and 7.

For Run III, given the levelling scheme, all kicks remain below $\unit[1]{\sigma}$ for the first two turns, and just reach it on the third turn for Beam~2, which is tolerable for the collimation system. However, in the worst case scenario, the kick crosses $\unit[1.1]{\sigma}$ on the second turn, which is above the $\unit[1]{MJ}$ limit ($\unit[1]{\sigma}$ for Run III beam parameters).

In HL-LHC, the kick on Beam~1 crosses the threshold, reaching up to $\unit[1.6]{\sigma}$ on the second turn for round optics. For flat optics, the kicks are significantly reduced due to the fact that the beta function is decreased in the crossing plane, which is also the kicking plane. Even after three turns, the orbit excursion would not reach the critical limit. Also, since this is an instantaneous kick that remains, and does not increase in strength over multiple turns, the orbit excursion would not change significantly from what is presented even if the beam remained in the machine for more than three turns (c.f. Fig.~\ref{fig:orbitBehavior}).

\subsection{Beam-Beam Kick Conclusions}
As one beam is extracted from the LHC, the sudden loss of the coherent Beam-Beam Kick (BBK) on the remaining beam can give a significant orbit distortion of up to $\unit[1.6]{\sigma}$ in the HL-LHC era. Delays between dumping the two beams of up to three turns have been observed on multiple occasions in the LHC. While this can be shortened following a redesign of the interlock and beam dumping system, it can at the minimum be one turn since the abort gaps of the two beams are not synchronized in the dump region, IR6.

For nominal LHC Run III scenarios, the kicks reach a maximum of $\unit[1]{\sigma}$, which is not critical. However, if one were to use the fully squeezed optics (\textit{end of SB}), with a larger than nominal bunch intensity of $\unit[1.5\times10^{11}]{p^+}$, the kick could reach above $\unit[1.1]{\sigma}$ on the second turn, or $\unit[1.2]{\sigma}$ on the third turn. While the losses resulting from this are not instantaneous due to the betatron motion of the beam, and also depend on the phase advances between the IPs and the primary collimators, it should be avoided without detailed loss studies.

In HL-LHC the results are similar; while the bunch intensity is larger than in Run III, the crossing angle is also increased. In round optics, an orbit excursion of $\unit[1.6]{\sigma}$ is reached on the second turn, breaching the $\unit[1.5]{\sigma}$ threshold. This should be avoided, and the machine protection system must ensure that there is no more than one turn delay between the two beam dumps. For Beam~2, this would mean that there are two turns of losses, however the $\unit[1.5]{\sigma}$ limit is not breached until turn three. 

For flat optics in the HL-LHC, the normalized kicks are smaller and fall well below the threshold.

The beams should always be linked for high-intensity operations, in both Run III and HL-LHC.

%\FloatBarrier

\section{Beam-Beam Compensating Wires}

\begin{figure}[!htb]
   \centering
   \includegraphics*[width=0.48\textwidth]{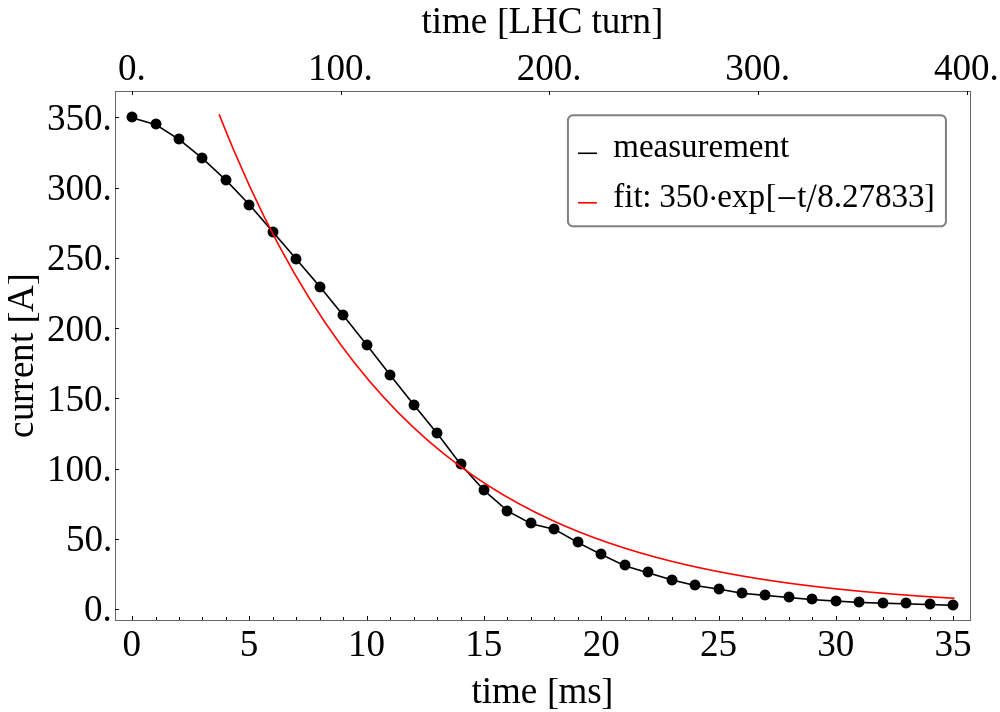}
   \caption{Measured current decay following an electrical short in the power supply to be used for the BBCW, including an exponential fit neglecting the initial part.}
   \label{fig:BBCWcurrentDecay}
\end{figure}

The Beam-Beam compensating wires (BBCW) are wires installed inside the beam pipe, parallel to the beam. They are to be connected to a power supply such that the current flows in the same direction as the beam, compensating for the beam-beam effects around the collision point~\cite{koutchouk_principle_nodate}. To analyze their impact on the beam from the perspective of machine protection, the wires are assumed to be installed in a free vacuum without any solid material in the vicinity. These assumptions are valid due to the long time constants on the order of milliseconds (see Fig.~\ref{fig:BBCWcurrentDecay}). Biot-Savart law is then used to calculate the field map, to which the multipolar decomposition is fitted. 

The wires are to be powered with a constant current. If a short were to occur, the current would approximately undergo an exponential decay. A measurement of this decay is seen in Fig.~\ref{fig:BBCWcurrentDecay}. As the current decays, the magnetic field produced by the wires weakens and the difference between the weakened field and the nominal field creates an effect on the beam. How this affects the beam depends on the setup and is discussed below.

The power supply in LHC Run III will be controlled by and interlocked using a \textit{function generator controller}~\cite{todd_radiation_nodate} in combination with a magnet interlock system, and a similar system is likely to be used in HL-LHC. The reaction time of these, to interlock on power cuts and to initiate a beam dump, was measured to $\unit[1.2]{ms}$.

\subsection{Run III}
In Run III, a pair of $\unit[1]{m}$ long wires will be installed in tertiary collimators right of IPs 1 and 5 (TCTPV.4R1.B2 and TCTPH.4R5.B2) for Beam~2. For Beam~1 they will be installed in tertiary collimators left of IPs 1 and 5, one per IP (TCTPV.4L1.B1 and TCTPH.4L5.B1)~\cite{BBCWrunIII_ECR,HLcollmeeting_BBCW_runIII}. The 'V' and the 'H' in the collimator names stand for "vertical" and "horizontal" and, thus, define the orientation of the wire pairs. The wires are embedded in the collimator jaws at a depth of $\unit[1]{mm}$. The collimator jaws will follow the nominal settings for tertiary collimators at $\unit[7.8]{\sigma}$ (for the smallest \textbeta\textsuperscript{*}), implying a distance of $\unit[11.2]{mm}$ horizontally or $\unit[9.9]{mm}$ vertically. The vertical case is used as the reference, due to its stronger effect on the beam. The polarity of the wires is such that the $\unit[350]{A}$ current in both of them flows in the beam direction, leading to even multipoles (quadrupolar, octupolar, ...) in the beam region. In case of a powering failure, the current decays in both wires simultaneously, leading to a change in the focusing gradient, inducing beta beating and a tune change. However, since there is no dipolar field component, the beam center should not be affected. 

While operating normally, their effect on the beam optics is compensated by quadrupole magnets. A power cut in the wires therefore leads to an overcompensation by the quadrupoles.

The quadrupolar field gradient is $\unit[1.4]{T/m}$, leading to a maximum absolute tune shift of 0.011 horizontally and -0.0075 vertically. Given that the fractional tunes are 0.31 horizontally and 0.32 vertically, there is no risk of hitting a low order resonance. The maximum beta beating is $7.5\%$ horizontally and $5\%$ vertically, which is within the general beta beating limit at $10\%$ in the LHC. However, this beating comes on top of unavoidable beta function errors already present in the machine. It is therefore important not to run the machine under these conditions and to issue a beam abort with a minimum delay.

With the present interlocking strategy and reference orbit it is however possible that the beams are offset by up to $\unit[2]{mm}$ per plane compared to the design orbit, making it possible for the beam to acquire an orbit excursion. Nevertheless, this is limited to less than $\unit[0.4]{\sigma}$ when the current has fully decayed after approximately $\unit[40]{ms}$, and is thus of little concern.

\subsection{HL-LHC}
In HL-LHC, the installation of the wires are not part of the baseline, but are under consideration. A preliminary design for the wires is that they are to be $\unit[3]{m}$ long, with a current of up to $\unit[150]{A}$~\cite{skoufaris_numerical_2019,HLcollmeeting}. There would only be one wire installed per location, upstream and downstream of both IPs 1 and 5, but the exact location is yet to be defined. The wires are most efficient when the ratio between the horizontal and the vertical beta functions are 2 or 0.5, implying they be installed right after the separation dipole, D1, or after the fourth quadrupole as seen from the IP, Q4~\cite{fartoukh_compensation_2015}, and a preliminary location after the Q4 magnets around IPs 1 and 5 ($\unit[\pm 195]{m}$) is reserved~\cite{HLcollmeeting}. 

The wires will be offset in the same direction as the crossing plane. They will be in the shadow of the nearby tertiary collimators, meaning that the distance to the beam will be $\unit[>10.4]{\sigma}$~\cite{skoufaris_numerical_2019,HLcollmeeting}. The magnetic field at the location of the beam is perpendicular to the offset, giving a kick in the direction of the wire. Since the square root of the beta function in the direction of the wire determines both the \textit{minimum} wire offset and the normalized kick on the beam, the \textit{maximum} normalized kick is independent of the beta function. The kick is thus calculated as follows, from one single wire:

Wire magnetic field:
\begin{equation}
B = \frac{\mu_0 I}{r [m]} = \frac{\mu_0 I}{r [\sigma] \cdot \sqrt{\epsilon_g \beta(s)}} 
\label{eq:2}
\end{equation}

where $\mu_0$ is the vacuum permeability, $I$ the wire current, $r$ the distance between the beam and the wire, i.e. the collimator setting. Inserting this magnetic field into Eq.~\ref{eq:kickIntegration}, the kick angle becomes:

\begin{equation}
\alpha \approx \frac{B l}{B\rho}\sqrt{\frac{\beta(s)}{\epsilon_g}} = \frac{\mu_0 I l}{B\rho \epsilon_g r[\sigma]}
\label{eq:3}
\end{equation}

\begin{figure}[!htb]
   \centering
   \includegraphics*[width=0.48\textwidth]{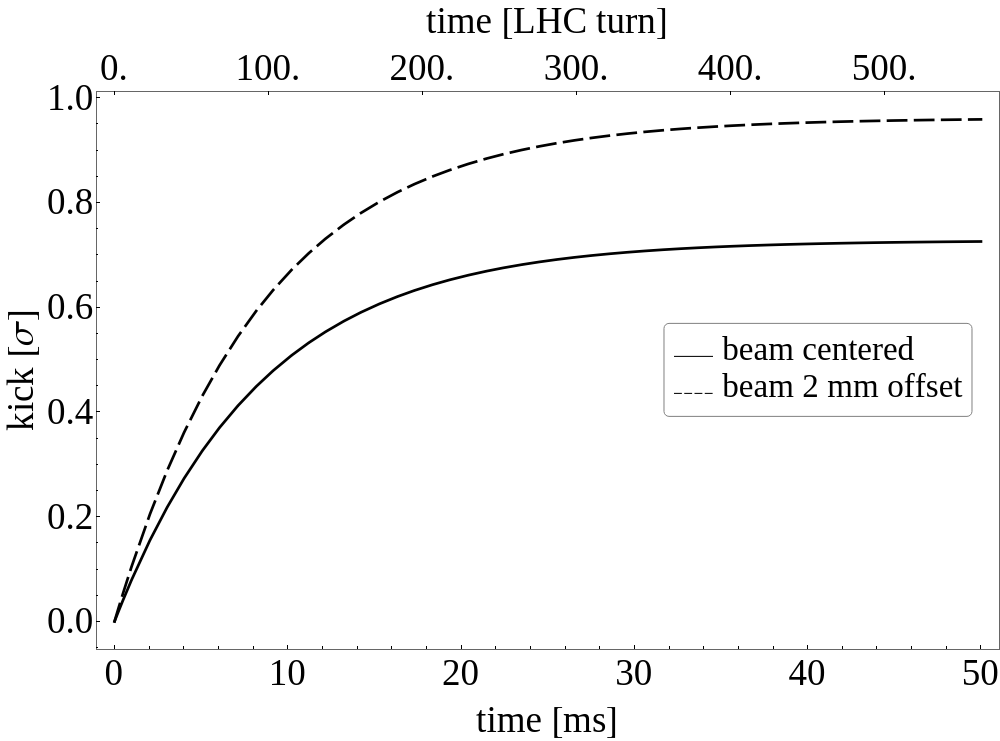}
   \caption{Kick on beam following a power failure of the wire. The power supply is assumed to be the same type as for Run III.}
   \label{fig:BBCWHLoffset}
\end{figure}

Since there is only one wire, a dipolar field component is created in the beam region. A powering failure of the wire would thus kick the beam. The result of this can be seen in Fig.~\ref{fig:BBCWHLoffset}. The kick reaches up to about $\unit[0.7]{\sigma}$ when the beam is initially centered, or $\unit[1]{\sigma}$ if it is initially offset by $\unit[2]{mm}$. The larger kick than for Run III is mainly due to the presence of the dipolar field, but there is also a slight increase due to the larger integrated current ($\unit[450]{Am}$ vs $\unit[350]{Am}$) and the tighter collimator jaw setting, allowing a smaller wire distance of $\unit[8.2]{mm}$. Note that due to the relatively slow onset of this kick, the beam will remain on a shifting closed orbit with an offset a factor $2\sin(\pi Q)\approx0.6$ smaller than the applied kick (c.f. Fig.~\ref{fig:orbitBehavior}).

\subsection{Conclusions}

The Beam-Beam Compensating Wires are not a major concern in either LHC Run III or HL-LHC. In Run III, due to the quadrupolar field, it can induce a beta beating up to $7.5\%$ and, at worst case, an orbit excursion up to $\unit[0.4]{\sigma}$. The beta beating comes on top of other sources of beating already present in the machine, such that the $10\%$ beta beating limit could be breached. The orbit excursion will also come on top of other sources, in particular the orbit excursion induced during the beam dump (c.f Section~\ref{BBK}). It is thus important to not run under such conditions, and to extract the beams as soon as a failure is detected, within $\unit[10]{ms}$ to provide a sufficient margin.

In HL-LHC, due to the dipolar field, the orbit excursion would be significantly worse, with the kick reaching almost $\unit[1]{\sigma}$ when the current is fully depleted. As during Run III, this orbit excursion would come on top of other sources, and a beam dump should thus be executed within $\unit[5]{ms}$ to provide sufficient margin when considering the loss of the Beam-Beam Kick.

Given the $\unit[1.2]{ms}$ reaction time of the interlock, there is a sufficient margin in both cases.

\section{LHC transverse damper - ADT}

The LHC transverse dampers, referred to as the ADT~\cite{gorbatchev_transverse_2001}, are a system of high resolution BPMs and electric kicker plates, that can damp transverse oscillations and is necessary for keeping the beam stable in the LHC~\cite{hofle_transverse_nodate}. Aside from damping the beam, the ADT can also excite it, both with a white noise mode resulting in a larger effective emittance, and also coherently, giving the beam center increasing oscillation amplitude around the closed orbit.

There are 4 pairs of kicker plates per beam, each $\unit[1.5]{m}$ long with a gap of $\unit[52]{mm}$~\cite{gorbatchev_transverse_2001}. They can work both on individual bunches and on the full beam at once. For single bunch resolution, the maximum effective voltage seen by the beam is $\unit[1]{kV}$, and for other modes it is $\unit[7.5]{kV}$. 

\subsection{Coherent Excitation}
\begin{figure}[!htb]
   \centering
   \includegraphics*[width=0.48\textwidth]{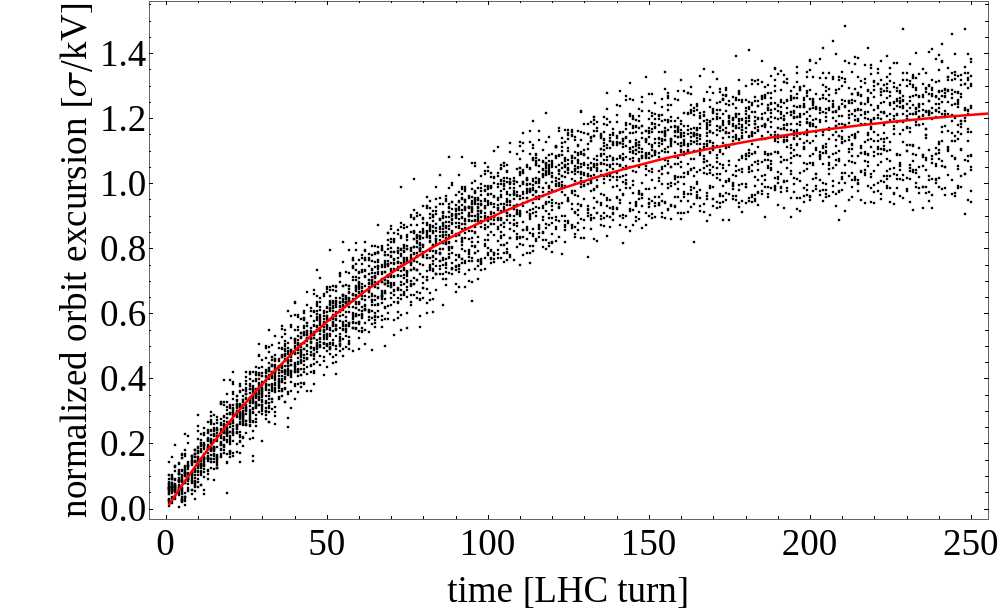}
   \caption{Measured coherent excitation of the beam by the ADT, normalized to the applied voltage. The beam energy was $\unit[6.5]{TeV}$, and the data consists of 26 separate measurement series of varying duration. Note that the orbit excursion ($\sigma$) in this plot is based on a normalized emittance of $\unit[3.5]{\mu m\cdot rad}$. Each point is the orbit excursion on a given turn in each of the 26 measurements.}
   \label{fig:ADToffset}
\end{figure}

Issues with the ADT can, in worst case, lead to unintentional coherent excitations of the beam as was observed during a test in 2018~\cite{ADTgainIssue}, and to quantify this a series of tests were run on the ADT using different modes of operations. The measured orbit excursion, normalized to the applied voltage, over time is shown in Fig.~\ref{fig:ADToffset}. 

During normal operations, the ADT will damp the beam in parallel to any excitations that it provides to the beam, since the kicker output is a superposition of the damping and the excitation. This works as an inherent safety measure, as the damping strength depends on the oscillation amplitude of the individual bunches; if the bunches are excited coherently with a static voltage, they will reach an equilibrium with the damping at a certain amplitude. This is the reason for the flattening of the curve in Fig.~\ref{fig:ADToffset}. 

Time-averaging over one period of betatron movement, the normalized orbit excursion over time, $\sigma(t)$ can be deduced from the following equation:

$$\frac{d}{dt}\sigma(t) = k-d\cdot \sigma(t)$$

where $k$ is the normalized kick per unit of time, and $d$ is the damping. The damping strength is proportional to the orbit excursion. Solving this equation and applying appropriate boundary conditions, gives the following expression:

$$\sigma(t) = \frac{k}{d}\big(1-\exp[-d\cdot t]\big)$$

Taking the time unit as turns, the damping strength $d$ can thus be interpreted as the damping time in units of turns. 

From the measurements, the fraction $k/d$ and the time constant $\tau$ were fitted to $\unit[1.274\pm0.004]{\sigma/kV}$ and $\unit[82.8\pm0.7]{turns}$ for a beam energy of $\unit[6.5]{TeV}$. At the injection energy of $\unit[450]{GeV}$, the parameters were fitted to $k/d=\unit[2.379\pm0.008]{\sigma/kV}$ and $\tau=\unit[44.5\pm0.4]{turns}$. Only the uncertainty of the fit is included. 

Since the kick is inversely proportional to the energy, and the normalization of the orbit excursion to the square root of the energy, $k$ is inversely proportional to the square root of the energy. Between $450$ and $\unit[6500]{GeV}$, there should thus be a factor $(44.5/82.8)\sqrt{6500/450} \approx 2.0$, whereas dividing the fitted $k/d$ values gives a factor $\approx 1.9$. This discrepancy could be from measurement errors, or due to the fact that the ADT samples the preceding four turns to determine the strength of the kick used for damping. However, for the purpose of this analysis, this error is deemed small enough.

\begin{figure}[!htb]
   \centering
   \includegraphics*[width=0.48\textwidth]{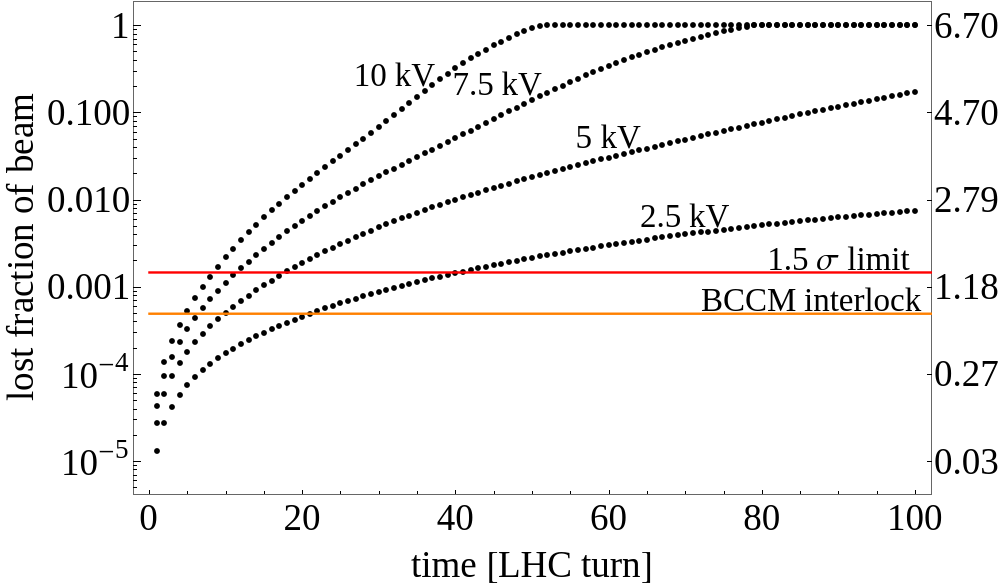}
   \caption{Estimate of the lost fraction of beam over time for ADT kicks with varying voltages and a beam energy of $\unit[7]{TeV}$. The red line indicates the $\unit[1.5]{\sigma}$ limit, whereas the orange line shows the interlock limit of the beam current change monitor (BCCM).}
   \label{fig:ADTresult}
\end{figure}

\subsection{Scaling to High Luminosity LHC}
From the fit of the results in Fig.~\ref{fig:ADToffset}, the induced losses over time for a given voltage can be estimated. This is shown in Fig.~\ref{fig:ADTresult} as an estimate on the lost fraction of beam over time, for a beam energy of $\unit[7]{TeV}$ and HL-LHC primary collimator gap settings ($\unit[6.7]{\sigma}$, c.f. Table~\ref{tab:collSettings}). The $\unit[1.5]{\sigma}$ limit is shown by the red line. For this scaling, it is assumed that the same damping time will be used as for the $\unit[6.5]{TeV}$ measurement above, and that the same beta functions at the ADT kickers apply.

An ADT voltage of a maximum of $\unit[10]{kV}$ was included to take potential upgrades for HL-LHC into account. This case constitutes the fastest possibility of going from dump threshold to the $\unit[1.5]{\sigma}$ limit, as the increase in losses is largest in the beginning of the excitation, for any voltage. If only a small part of the beam is excited, larger orbit excursions would be needed to reach both the dump threshold and the $\unit[1.5]{\sigma}$ limit. This, together with the stronger damping at larger orbit excursions, means that the loss increase would be slower than if the full beam is excited.

Taking the interlock threshold of the beam current change monitor~\cite{BCCM_EDMS}, at $3\unit[\times10^{11}]{p^+}$, or a fraction of $5\times10^{-4}$ of the full beam, there would thus be a margin of four turns to dump the beams in the worst case, which is tolerable. In parallel to this, there are also BLMs that could react on the beam losses. The BLMs react even faster since they measure the beam losses directly, whereas the BCCM can only measure a change in the total beam intensity.

\subsection{ADT Conclusions}

While issues with the ADT are likely to be due to human error, which can be protected against by software and procedure, a malfunction leading to beam losses and a beam dump has been observed. For coherent excitations, the ADT can be fast, exciting the beam to the orbit excursion limit in 10 turns at the maximum voltage of $\unit[7.5]{kV}$. This is slow enough for the BLM system as well as the new beam current change monitor to detect the losses and extract the beams. However, there is not much margin for increasing the power of the kickers. If the hollow electron lens is employed, it must be ensured that the ADT acts the same way on the untouched witness bunches as it does on other bunches in the beam, or it could effectively risk circumventing the machine protection systems.

\section{Symmetric and quickly propagating Triplet Quench}
\begin{figure}[!htb]
   \centering
   \includegraphics*[width=0.48\textwidth]{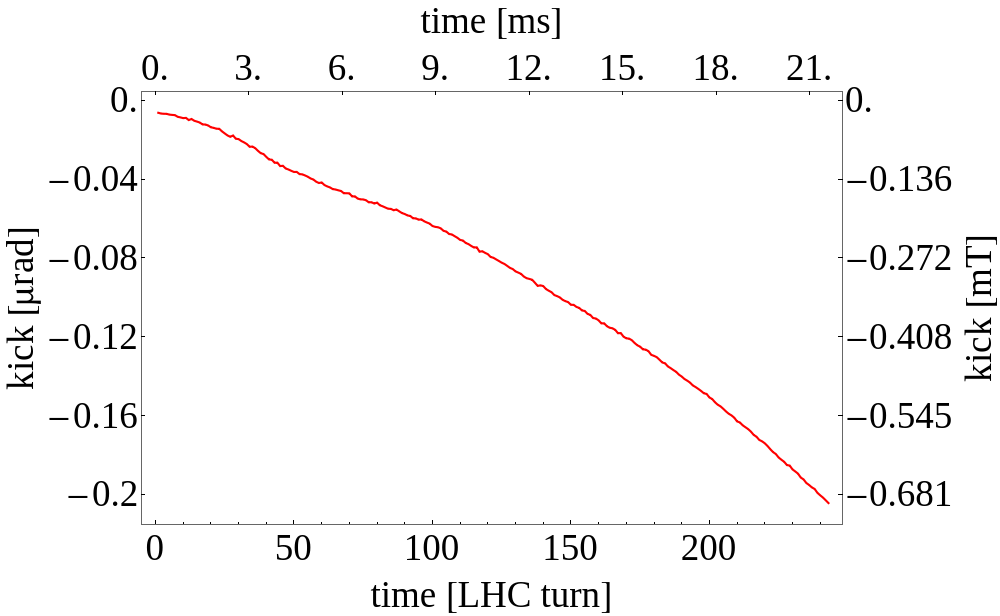}
   \caption{Kick on the beam during a quench of one of the ATLAS Q1 magnets, calculated from the measured BPM response.}
   \label{fig:tripletQuench}
\end{figure}

Having the largest beta functions in the whole ring, the final focusing triplet magnets around IP1 and IP5 are among the most critical elements. As the magnets are superconducting, powering failures are in general not considered critical due to the long time constants involved. However, quenches featuring unusually fast normal-zone propagation can produce significant kicks on short timescales, as was observed for the ATLAS Q1 quadrupole right of the IP on June 3rd 2018~\cite{RQX1report}. In this event, it is believed that a symmetric quench occurred in the top part of the magnet, which was, due to its symmetry, not detected by the quench detection system until after $\unit[40.5]{ms}$~\cite{QPS_EDMS,felixQPS,denzQPS}. Meanwhile, an orbit distortion developed, resulting in a beam dump due to losses $\unit[21.5]{ms}$ after the start of the quench. 

While changes to quadrupole gradients normally only lead to beta beating, in the triplets the beams are offset by up to $\unit[\sim6]{mm}$ to enable the crossing, leading to a distortion of the bunch center-of-mass orbit as well as beta beating.

The vertical kick, $\Delta y'$, was determined by fitting the following expression for a changing closed orbit to beam position monitors (BPMs) distributed throughout the LHC:

$$\frac{y(s)}{\sqrt{\beta(s)}} = \Delta y' \sqrt{\beta_0} \frac{\cos[|\varphi(s)-\varphi_0|-\pi Q_y]}{2\sin[\pi Q_y]}$$

where $s$ is the position of the BPMs in the ring, $y(s)$ the measured vertical positions at the BPMs, $\beta(s)$ the beta function at the BPMs, $\beta_0$ the average beta function in the magnet that quenched, $\varphi(s)$ the phase advance at the BPMs, $\varphi_0$ the average phase advance in the magnet, and $Q_y$ the vertical tune. The fitted kick is shown in Fig.~\ref{fig:tripletQuench}, both in radians and in the corresponding average magnetic field over the length of the magnet seen by the beam. The beam was dumped 242 turns after the start of the quench, with the amplitude of the kick being $\unit[0.2]{\mu rad}=\unit[0.46]{\sigma}$.

In this event, there were three unexpected circumstances:
\begin{itemize}
    \item The quench propagation speed was determined by fitting circuit simulations to the measurements of the current, giving a longitudinal propagation speed of about $\unit[50]{m/s}$ along the superconducting cable, and a turn-to-turn propagation time of $\unit[1]{ms}$.
    \item The current drop of $\Delta I = \unit[1.69]{A}$ at the time of the beam dump is not in itself enough to explain the resulting kick of $\unit[0.2]{\mu rad}$.
    \item Only Beam~1 was affected measurably by the quench.
\end{itemize}

The abnormally high quench propagation speed, about two to three times faster than expected under nominal circumstances, is believed to have been due to the bath temperature being at $\unit[2.16]{K}$. This is close to the phase transition point for superfluid Helium, rather than at the nominal $\unit[1.9]{K}$. Collision debris from the IP1 collision point could then have triggered a symmetric quench in the top part of the magnet, which together with the increased bath temperature caused the quickly propagating quench.

As for the second and third points, the magnetic field at the position of the beam, having a vertical offset and negligible horizontal offset, is
$$\Delta B_x(x=0,y) = T\Delta I y $$

where $T\approx \unit[30.31]{T/m/kA}$ is the magnet transfer function under static conditions at $I=\unit[6.2]{kA}$, the magnet current at the time of the quench~\cite{RQX1report}. The kick is then calculated using the average beam orbit $y$ in the magnet, the magnetic length $l$ and the magnetic rigidity of the beam $B\rho$:

$$\Delta y' =  T\Delta I y \frac{l}{B\rho} \approx \unit[0.068]{\mu rad}$$

This is about three times smaller than the kick deduced from the measurements in Fig.~\ref{fig:tripletQuench}. Furthermore, repeating the calculations for Beam~2 gives a kick of $\unit[0.060]{\mu rad}$, similar to that of Beam~1, yet no orbit distortion was measured. A possible explanation for this was found by assuming that the transport current in the quenched conductor redistributed non-uniformly in the strands of the cable due to inhomogenous magneto-resistivity. Then, the net field change for Beam~2, which is in the bottom part of the magnet, is close to zero, while the net field change for Beam~1 at the top of the magnet, is boosted. More details on these simulations are summarized in a technical note~\cite{RQX1report}.

\subsection{Scaling to HL-LHC}
In the HL-LHC the beta functions in the triplet quadrupoles will be increased significantly and the beams will be offset further, up to $\unit[\sim17]{mm}$, increasing the criticality of an undetected quickly propagating quench. In order to study this, the magnet simulations of the quench described above have been adjusted to the HL-LHC triplet magnets, with otherwise similar assumptions on the propagation speed of the quench. The current redistribution was neglected as it is a feature that has not been verified.

\begin{figure}[!htb]
   \centering
   \includegraphics*[width=0.48\textwidth]{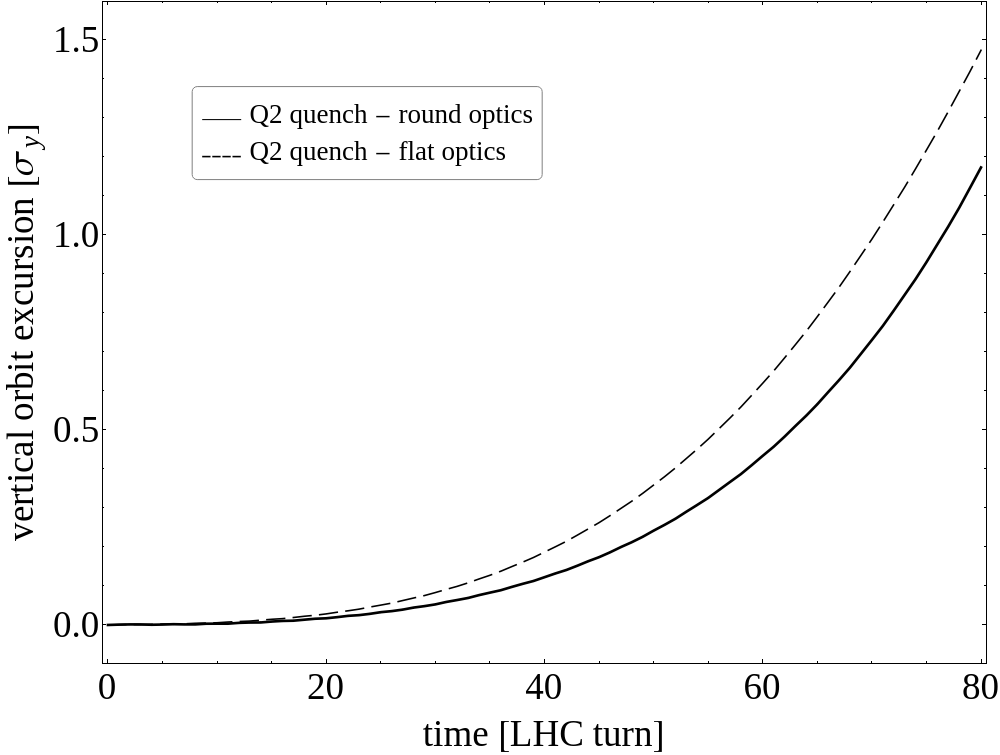}
   \caption{Orbit excursion of the beam following a kick due to a quench in one of the Q2 magnets. The solid line is for round optics, while the dashed line is for flat optics.}
   \label{fig:tripletQuenchSim}
\end{figure}

The resulting normalized kick is shown in Fig.~\ref{fig:tripletQuenchSim}. The plot shows the results for round and flat optics. Flat optics constitute the worst case, since the magnetic field is parallel to the crossing plane and the beta function is increased in the perpendicular plane in flat optics as compared to round optics. The orbit excursion on turn 80 is $\unit[1.5]{\sigma}$ in the flat case. 

The calculated kick is significantly stronger than the one measured (Fig.~\ref{fig:tripletQuench}) and the critical orbit excursion is reached around 80 turns after the start of the quench, making it a fast failure. A symmetric quench detection is envisioned for HL-LHC~\cite{HLTDRv01_chapter7}, but due to a quench evaluation time of up to $\unit[10]{ms}\approx\unit[90]{turns}$, the beam abort would likely be triggered by beam losses in the collimation region, measured by the BLMs or the beam current change monitor. With an interlock limit of $3\times10^{11}$ protons lost from the beam, the beam current change monitor would trigger a dump at $\sim\unit[0.8]{\sigma}$. The interlock would thus be breached on turn 65, giving some margin before the $\unit[1.5]{\sigma}$ limit is reached.

Furthermore, due to the lack of a validated model, the current redistribution which is believed to explain the large kick in the measured quench has not been included. The current redistribution in the measured case caused roughly a factor three larger kick than otherwise expected, and presently it cannot be excluded that a similar effect would be present in the HL-LHC. Scaling the kick by the same factor, the detection and dump margin would be about $\unit[10]{turns}$ and is thus not considered to be a concern.

\subsection{Symmetric Triplet Quench Conclusions}
A fast symmetric quench of one of the ATLAS triplet quadrupoles has been observed with beam in the machine. Due to an abnormally high cryo temperature of $\unit[2.16]{K}$, the quench propagation speed was estimated at $\unit[50]{m/s}$, leading to a significant current drop and beam orbit excursion. The beam was dumped on beam losses in the collimation region after 242 turns. Replicating this quench propagation speed with HL-LHC optics, the effect is faster, reaching the $\unit[1.5]{\sigma}$ limit in 80 turns, from the onset of the quench. Furthermore, adding the effect of the current redistribution around the normal resistive zone similarly to the observed quench, the limit could be reached even faster, in around 55 turns.

While this is significantly faster than previously expected, it is still within the margins for detection and dump trigger by the BLMs and the new beam current change monitor. Effects of superconducting magnet quenches on the beam should however be reiterated if there are fundamental changes to beam optics or interlock thresholds. In particular, it must be ensured that this holds true for use with the hollow electron lens. An interlock on the helium bath temperature, triggering a ramp down of the magnet, could prevent this kind of fast-propagating quenches.
%\FloatBarrier

\section{Conclusions}

The large stored energy in both the nominal LHC and the future HL-LHC, makes it vital to ensure that uncontrolled beam losses are kept minimal or the machine could be damaged. The goal of this paper was to analyze \textit{fast failures}, that is, failures that risk causing damaging beam losses within $\unit[10]{ms}$. This paper covers a broad range of failure scenarios, and recommendations for protection strategies have been proposed for the critical cases.

Failures of magnet protection systems constitutes by far the most critical scenarios for machine protection. It is required that a QH or CLIQ unit by design cannot fire before the beams are extracted. Nevertheless, spurious discharges of single units have been observed for QHs in the past and cannot be excluded. CLIQ can potentially lead to an orbit excursion of $\unit[3]{\sigma}$ within one turn, rising to more than $\unit[10]{\sigma}$ before the beams could be dumped. This is an unacceptable risk and a new connection scheme was proposed to mitigate this. It is required that any CLIQ unit is connected symmetrically around the magnet poles, such that no significant dipolar field components are produced. This reduces the orbit excursion to $\unit[1.5]{\sigma}$ in 17 turns, providing sufficient margin for detecting and interlocking on any spurious discharges.
The main magnetic field component being skew octupolar in this case, it is important that the beam orbit is measured in detail and kept to within $\unit[2]{mm}$ of the design value for the baseline round optics, or within $\unit[1]{mm}$ for flat optics. Any optics changes are to be evaluated for their effects on this orbit and increases in the orbit offset must be avoided without a detailed loss study. 

The QHs in the new HL-LHC magnets are optimized to produce minimal dipolar fields when all circuits fire. For the triplets, a strong dipolar field is however still produced. In the Q2 triplets the kick is up to $\unit[8.2]{\sigma}$, reiterating the need to extract the beams before triggering the quench protection systems.

For spurious discharges of single QH circuits, the worst kicks are $\unit[1.78]{\sigma}$ and $\unit[1.38]{\sigma}$ in the D1 separation dipole, for flat and round optics respectively. In the Q2, the kicks are similar. Detection of spurious QH discharges in the most critical magnets, D1, Q1, Q2 and Q3, is thus required. For flat optics, a detailed loss study is necessary and the necessity of a phase advance interlock between D1/Q2 and the primary collimators should be evaluated.

A loss of the Beam-Beam Kick is inevitable in the LHC as there is a minimum one turn delay between dumping of the two beams, where the orbit of the remaining beam is disturbed. Currently, this delay can be up to three turns long. Due to different optics and a higher bunch intensity in LHC Run III and HL-LHC, the resulting orbit excursion of the remaining beam is increased. For HL-LHC, the orbit excursion can reach up $\unit[1.6]{\sigma}$ on the second turn, which can add up with any other orbit excursions already present. It is thus required that the machine protection system is adjusted such that both beams are dumped within one turn of each other.

In HL-LHC, a coherent ADT excitation can cause a $\unit[1.5]{\sigma}$ orbit excursion in ten turns. This can be protected against via beam losses. However, in case the hollow electron lens is employed, it must be ensured that the ADT cannot circumvent the protection of the machine by only acting on normal bunches while ignoring the witness bunches.

Detection of beam losses are a sufficient protection for triplet quenches, whereas failures with the Beam-Beam Compensating Wires can be interlocked at the power supply.

In summary, all known critical failures analyzed in this paper have been addressed and mitigated. 

\section{Acknowledgements}
We would like to express our gratitude to the LHC operations team and Daniel Valuch (CERN) of the ADT team for collaborating on the experiments that are referred to in this paper. We would also like to thank Roderik Bruce (CERN) for insightful discussions, and Laurent Ceccone and Richard Mompo (CERN) for performing measurements on interlock delays.

\bibliography{main}

%\clearpage
\appendix
\section{Optics}

\begin{table*}[!hb]
   \centering
   \caption{Main optics parameters for the different triplet magnets, in \textit{HLLHCv1.4} optics, round collision with \textbeta\textsuperscript{*} of $\unit[15]{cm}$. Since the parameters change throughout the magnets, the average over the magnet is shown.}
   \begin{tabular}{lccccc}
        Magnet & Length [m] & Hor. beta [m] & Ver. beta [m] & Hor. orbit [mm] & Ver. orbit [mm] \\
        \toprule
        \textbf{Q1} \\
		MQXFA.B1L1      & 4.2 & 7580.5 & 4634.8 & -9.9 & 0 \\
		MQXFA.A1L1      & 4.2 & 4421.0 & 3989.8 & -7.6 & 0 \\
		MQXFA.A1R1      & 4.2 & 3982.3 & 4430.0 &  7.2 & 0 \\
		MQXFA.B1R1      & 4.2 & 4626.1 & 7594.4 &  7.8 & 0 \\
		MQXFA.B1L5      & 4.2 & 7618.3 & 4631.0 &    0 & -7.8 \\
		MQXFA.A1L5      & 4.2 & 4444.0 & 3986.5 &    0 & -7.2 \\
		MQXFA.A1R5      & 4.2 & 4002.3 & 4426.4 &    0 &  7.6 \\
		MQXFA.B1R5      & 4.2 & 4649.3 & 7588.2 &    0 &  9.9 \\
		\hline
		\textbf{Q2} \\
		MQXFB.B2L1  & 7.15 & 19371.3 &  8307.6 & -16.0 & 0 \\
		MQXFB.A2L1  & 7.15 & 17671.6 &  4804.3 & -15.2 & 0 \\
		MQXFB.A2R1  & 7.15 &  4795.4 & 17704.2 &   8.0 & 0 \\
		MQXFB.B2R1  & 7.15 &  8292.2 & 19407.1 &  10.5 & 0 \\
		MQXFB.B2L5  & 7.15 & 19468.1 &  8300.8 &     0 & -10.5 \\
		MQXFB.A2L5  & 7.15 & 17759.9 &  4800.4 &     0 &  -8.0 \\
		MQXFB.A2R5  & 7.15 &  4819.4 & 17689.8 &     0 &  15.2 \\
		MQXFB.B2R5  & 7.15 &  8333.7 & 19391.2 &     0 &  16.0 \\
        \hline
        \textbf{Q3} \\
        MQXFA.B3L1  & 4.2 &   8737.8 & 20990.2 & -11.0 & 0 \\
        MQXFA.A3L1  & 4.2 &  11213.5 & 17753.4 & -12.3 & 0 \\
        MQXFA.A3R1  & 4.2 &  17720.4 & 11234.2 &  15.5 & 0 \\
        MQXFA.B3R1  & 4.2 &  20951.2 &  8754.0 &  17.0 & 0 \\
        MQXFA.B3L5  & 4.2 &   8781.5 & 20973.0 &     0 & -17.0 \\
        MQXFA.A3L5  & 4.2 &  11269.5 & 17738.9 &     0 & -15.5 \\
        MQXFA.A3R5  & 4.2 &  17809.1 & 11225.1 &     0 &  12.3 \\
        MQXFA.B3R5  & 4.2 &  21056.1 &  8746.8 &     0 &  11.0 \\
        \bottomrule   
   \end{tabular}
   \label{tab:tripletOptics}
\end{table*}

\end{document}